%% file: EDCAadmissioncontrol.tex
\documentclass[journal, onecolumn, 12pt]{IEEEtran}

\usepackage{graphics}
\usepackage{rotating}
\usepackage{amsfonts}
\usepackage{amsmath}
\usepackage{subfigure}
\usepackage{verbatim}
\usepackage{enumerate}
\usepackage{setspace}
\usepackage[noadjust]{cite}
\usepackage{multirow}

\doublespace
\centerfigcaptionstrue

\usepackage{epsfig}
\graphicspath{{figures/}} \DeclareGraphicsExtensions{.eps,.ps}
\renewcommand{\thefootnote}{\fnsymbol{footnote}}

\begin{document}

\title{Multimedia Capacity Analysis of the IEEE 802.11e Contention-based Infrastructure Basic Service Set \footnotemark{$^{\dag}$}}

\author{\singlespace \normalsize \authorblockN{Inanc Inan \textit{Student Member}, Feyza Keceli \textit{Student Member}, and Ender Ayanoglu
\textit{Fellow}}\\
\authorblockA{Center for Pervasive Communications and Computing \\
Department of Electrical Engineering and Computer Science\\
The Henry Samueli School of Engineering\\
University of California, Irvine, 92697-2625\\
Email: \{iinan, fkeceli, ayanoglu\}@uci.edu}}

\maketitle

\footnotetext{$^{\dag}$ This work is supported by the Center for
Pervasive Communications and Computing, and by Natural Science
Foundation under Grant No. 0434928. Any opinions, findings, and
conclusions or recommendations expressed in this material are
those of authors and do not necessarily reflect the view of the
National Science Foundation.}

\renewcommand{\thefootnote}{\arabic{footnote}}

\input{abstract}
\input{introduction}
\input{EDCAoverview}
\input{related}
\input{cycletime}
\input{validation}
\input{admission}
\input{conclusion}


\bibliographystyle{IEEEtran}
\bibliography{IEEEabrv,C:/INANCINAN/bibliography/standards,C:/INANCINAN/bibliography/HCCA,C:/INANCINAN/bibliography/simulations,C:/INANCINAN/bibliography/channel,C:/INANCINAN/bibliography/books,C:/INANCINAN/bibliography/EDCAanalysis,C:/INANCINAN/bibliography/mypapers,C:/INANCINAN/bibliography/myreports}

\input{figures}

\end{document}

%% file: abstract.tex
\begin{abstract}
We first propose a simple mathematical analysis framework for the
Enhanced Distributed Channel Access (EDCA) function of the
recently ratified IEEE 802.11e standard. Our analysis considers
the fact that the distributed random access systems exhibit cyclic
behavior. The proposed model is valid for arbitrary assignments of
AC-specific Arbitration Interframe Space (AIFS) values and
Contention Window (CW) sizes and is the first that considers an
arbitrary distribution of active Access Categories (ACs) at the
stations. Validating the theoretical results via extensive
simulations, we show that the proposed analysis accurately
captures the EDCA saturation performance.
Next, we propose a framework for multimedia capacity analysis of
the EDCA function. We calculate an accurate station- and
AC-specific queue utilization ratio by appropriately weighing the
service time predictions of the cycle time model for different
number of active stations. Based on the calculated queue
utilization ratio, we design a simple model-based admission
control scheme. We show that the proposed call admission control
algorithm maintains satisfactory user-perceived quality for
coexisting voice and video connections in an infrastructure BSS
and does not present over- or under-admission problems of
previously proposed models in the literature.
\end{abstract}

%% file: introduction.tex
\section{Introduction} \label{sec:introduction}

The IEEE 802.11 standard \cite{802.11} defines the Distributed
Coordination Function (DCF) which provides best-effort service at
the Medium Access Control (MAC) layer of the Wireless Local Area
Networks (WLANs). The recently ratified IEEE 802.11e standard
\cite{802.11e} specifies the Hybrid Coordination Function (HCF)
which enables prioritized and parameterized Quality-of-Service
(QoS) services at the MAC layer, on top of DCF. The HCF combines a
distributed contention-based channel access mechanism, referred to
as Enhanced Distributed Channel Access (EDCA), and a centralized
polling-based channel access mechanism, referred to as HCF
Controlled Channel Access (HCCA). In this paper, we confine our
analysis to the EDCA scheme, which uses Carrier Sense Multiple
Access with Collision Avoidance (CSMA/CA) and slotted Binary
Exponential Backoff (BEB) mechanism as the basic access method.
The EDCA defines multiple Access Categories (AC) with AC-specific
Contention Window (CW) sizes, Arbitration Interframe Space (AIFS)
values, and Transmit Opportunity (TXOP) limits to support
MAC-level QoS \cite{802.11e}.

We first propose a simple analytical model in order to assess the
performance of EDCA function accurately for the saturation
(asymptotic) case when each contending AC always has a packet in
service (each AC is always active). The analysis of the saturation
provides in-depth understanding and insights into the random
access schemes and the effect of different contention parameters
on the performance. Moreover, as we will show, the saturation
figures can effectively be used in network capacity estimation.
Our analysis is based on the fact that a random access system
exhibits cyclic behavior. A cycle time is defined as the duration
in which an arbitrary tagged user successfully transmits one
packet on average \cite{Medepalli05}. We derive the explicit
mathematical expression of the station- and AC-specific EDCA cycle
time. The derivation considers the AIFS and CW differentiation by
employing a simple average collision probability analysis. Our
formulation is also the first to consider the scenario such that
the number of active ACs may vary from station to station. As a
direct result, the proposed model also takes the internal
collisions into account in the case of a station having more than
one active ACs. We use the EDCA cycle time to predict the average
throughput and the average service time in saturation. We show
that the results obtained using the cycle time model closely
follow the accurate predictions of the previously proposed more
complex analytical models and simulation results. Our cycle time
analysis can serve as a simple and practical alternative model for
EDCA saturation performance analysis.

Due to its contention-based nature, EDCA cannot provide
parameterized QoS for realtime applications that require strict
QoS guarantees, if the network load and parameters are not tuned
such that the network is operating in nonsaturated state
\cite{XChen06},\cite{Zhai05}. Although the use of an admission
control algorithm is recommended in \cite{802.11e} to limit the
network load for QoS provision, no algorithm is specified. A loose
capacity estimation is harmful for admission control, since the
quality of ongoing flows will be jeopardized. Conversely, an
underestimation of the network capacity results in fewer number of
admitted flows than the network can support.

Rather than designing a new and complex access model with a large
number of states in order to calculate the EDCA capacity (in
nonsaturation), we propose a novel, simple, and accurate framework
which directly employs the results of the proposed simple
saturation analysis. An approximate station- and AC-specific
average service time is calculated by weighing the average service
time calculated using cycle time model for different number of
active stations. Given the average station- and AC-specific
traffic load, the average service time is directly translated into
a station- and AC-specific queue utilization ratio (note that
since all the measures are station- and AC-specific, the proposed
framework considers the potential unbalanced traffic load in the
802.11e BSS uplink and downlink). Next, we design a novel
centralized EDCA admission control algorithm the admission
decisions of which are based on the queue utilization ratio. The
key point is that the delay guarantee of realtime applications is
only possible when the queue utilization ratio of active
multimedia flows is smaller than 1 (i.e., when the MAC queue is
stable). Comparing the theoretical results with simulations, we
show that the proposed call admission control algorithm maintains
satisfactory user-perceived quality for coexisting voice and video
connections in an infrastructure BSS by limiting the maximum
number of admitted flows of each multimedia traffic type.
Comparison with extensive simulation results also reveals that the
proposed analysis does not result in an overestimation or a
significant underestimation of the network capacity. Another
attractive feature of the proposed scheme is that it fully
complies with the 802.11e standard.

The main contributions of this paper are three-fold; \textit{i)} a
simple average cycle time model to evaluate the performance of the
EDCA function in saturation for an arbitrary assignment of
AC-specific AIFS and CW values and an arbitrary distribution of
active ACs at the stations, \textit{ii)} an approximate capacity
estimation framework which weighs the saturation service times in
order to calculate the nonsaturation service time, and
\textit{iii)} a practical model-based admission control algorithm
to limit the number of admitted realtime multimedia flows in the
802.11e infrastructure BSS.

%% file: EDCAoverview.tex
\section{EDCA Overview}\label{sec:EDCAoverview}

The IEEE 802.11e EDCA is a QoS extension of IEEE 802.11 DCF. The
major enhancement to support QoS is that EDCA differentiates
packets using different priorities and maps them to specific ACs
that are buffered in separate queues at a station. Each AC$_{i}$
within a station ($0\leq i\leq i_{max}$, $i_{max}=3$ in
\cite{802.11e}) having its own EDCA parameters contends for the
channel independently of the others. Following the convention of
\cite{802.11e}, the larger the index $i$ is, the higher the
priority of the AC is. Levels of services are provided through
different assignments of the AC specific EDCA parameters; AIFS,
CW, and TXOP limits.

If there is a packet ready for transmission in the MAC queue of an
AC, the EDCA function must sense the channel to be idle for a
complete AIFS before it can start the transmission. The AIFS of an
AC is determined by using the MAC Information Base (MIB)
parameters as
\begin{align}
AIFS = SIFS + AIFSN \times T_{slot},
\end{align}

\noindent where $AIFSN$ is the AC-specific AIFS number, $SIFS$ is
the length of the Short Interframe Space and $T_{slot}$ is the
duration of a time slot.

If the channel is idle when the first packet arrives at the AC
queue, the packet can be directly transmitted as soon as the
channel is sensed to be idle for AIFS. Otherwise, a backoff
procedure is completed following the completion of AIFS before the
transmission of this packet. A uniformly distributed random
integer, namely a backoff value, is selected from the range
$[0,W]$.
The backoff counter is decremented at the slot boundary if the
previous time slot is idle. Should the channel be sensed busy at
any time slot during AIFS or backoff, the backoff procedure is
suspended at the current backoff value. The backoff resumes as
soon as the channel is sensed to be idle for AIFS again. When the
backoff counter reaches zero, the packet is transmitted in the
following slot.

The value of $W$ depends on the number of retransmissions the
current packet experienced. The initial value of $W$ is set to the
AC-specific $CW_{min}$. If the transmitter cannot receive an
Acknowledgment (ACK) packet from the receiver in a timeout
interval, the transmission is labeled as unsuccessful and the
packet is scheduled for retransmission. At each unsuccessful
transmission, the value of $W$ is doubled until the maximum
AC-specific $CW_{max}$ limit is reached. The value of $W$ is reset
to the AC-specific $CW_{min}$ if the transmission is successful,
or the retry limit is reached thus the packet is dropped.

The higher priority ACs are assigned smaller AIFSN. Therefore, the
higher priority ACs can either transmit or decrement their backoff
counters while lower priority ACs are still waiting in AIFS. This
results in higher priority ACs enjoying a relatively faster
progress through backoff slots. Moreover, the ACs with higher
priority may select backoff values from a comparably smaller CW
range. This approach prioritizes the access since a smaller CW
value means a smaller backoff delay before the transmission.

Upon gaining the access to the medium, each AC may carry out
multiple frame exchange sequences as long as the total access
duration does not go over a TXOP limit. Within a TXOP, the
transmissions are separated by SIFS. Multiple frame transmissions
in a TXOP can reduce the overhead due to contention. A TXOP limit
of zero corresponds to only one frame exchange per access.

An internal (virtual) collision within a station is handled by
granting the access to the AC with the highest priority. The ACs
with lower priority that suffer from a virtual collision run the
collision procedure as if an outside collision has occured
\cite{802.11e}.

%% file: related.tex
\section{Related Work}

In this section, we provide a brief summary of the studies in the
literature that are related to this work.

\subsection{Performance Analysis of EDCA in Saturation}

Three major saturation performance models have been proposed for
DCF; \textit{i)} assuming constant collision probability for each
station, Bianchi \cite{Bianchi00} developed a simple Discrete-Time
Markov Chain (DTMC) and the saturation throughput is obtained by
applying regenerative analysis to a generic slot time,
\textit{ii)} Cali \textit{et al.} \cite{Cali00} employed renewal
theory to analyze a \textit{p}-persistent variant of DCF with
persistence factor \textit{p} derived from the CW, and
\textit{iii)} Tay \textit{et al.} \cite{Tay01} instead used an
average value mathematical method to model DCF backoff procedure
and to calculate the average number of interruptions that the
backoff timer experiences. Having the common assumption of slot
homogeneity (for an arbitrary station, constant collision or
transmission probability at an arbitrary slot), these models
define all different renewal cycles all of which lead to accurate
saturation performance analysis. These major methods (especially
\cite{Bianchi00}) are modified by several researchers to include
an accurate treatment of the QoS features of the EDCA function
(AIFS and CW differentiation among ACs) in the saturation analysis
\cite{Xiao05,Kong04,Robinson04,Hui05,Zhu05,Inan07_ICC,Tao06,Zhao02,Banchs06,Lin06,Kuo03}.

Our approach in this paper is based on the observation that the
transmission behavior in the contention-based 802.11 WLAN follows
a pattern of periodic cycles. Previously, Medepalli \textit{et
al.} \cite{Medepalli05} provided explicit expressions for average
DCF cycle time and system throughput. Similarly, Kuo \textit{et
al.} \cite{Kuo03} calculated the EDCA transmission cycle assuming
equal collision probability for any AC. On the other hand, such an
assumption is shown to lead to analytical inaccuracies
\cite{Xiao05,Kong04,Robinson04,Hui05,Zhu05,Inan07_ICC,Tao06,Zhao02,Banchs06,Lin06}.
One of the main contributions of this paper is that we incorporate
accurate AIFS and CW differentiation calculation in the EDCA cycle
time analysis. We show that the cyclic behavior is observed on a
per AC per station basis in the EDCA. To maintain the simplicity
of the cycle time analysis, we employ averaging on the AC- and
station-specific collision probability. Another key contribution
of the proposed model is that our analysis is the first analytical
EDCA model to consider the possibility of the number of active ACs
varying from station to station. The comparison with more complex
and detailed theoretical and simulation models reveals that the
analytical accuracy is preserved when average cyclic time analysis
is used.

\subsection{Capacity Analysis and Admission Control in EDCA}

The Markov analysis of \cite{Bianchi00} is also modified by
several researchers to include the capacity analysis of the DCF or
EDCA function in nonsaturation
\cite{Duffy05,Shabdiz06,Cantieni05,Engelstad06}. A number of
queueing models have also been proposed to analyze delay
performance of a station or an AC under the assumption that the
traffic is uniformly distributed
\cite{XChen06,Zhai05,Tickoo04_2,Zhai04}. Some other queueing
models also assumed a MAC queue size of one packet to define a
Markovian framework for performance analysis
\cite{Foh02,Tantra06}.

There are also studies on capacity analysis and admission control
considering the infrastructure BSS where the AP usually has a
higher load in the downlink than the stations serving traffic in
the uplink. A group of studies mainly concentrated on capacity
analysis of only Voice-over-IP traffic for DCF and did not
consider traffic differentiation
\cite{Hui06,Medepalli05_2,Hole04,Garg03_1,Garg03_2,Cai06}. Gao
\textit{et al.} \cite{Gao06} and Cheng \textit{et al.}
\cite{Cheng06} calculated VoIP capacity of the WLAN when CW
differentiation among uplink and downlink flows are used. Another
group of studies defines parameter adaptation algorithms for QoS
enhancement and defines measurement-assisted call admission
control algorithms
\cite{Zhang04,Pong03,He05,Xiao04_gdpc1,Xiao04_gdpc2,Xiao05_gdpc3,Xiao06_gdpc}.

Being a very simple extension of the proposed cycle time analysis,
our approach in this paper provides an accurate multimedia traffic
capacity estimation in the case of traffic differentiation between
voice, video, and best-effort flows in the WLAN. Our analysis
considers the unbalanced traffic between the AP and the stations.
Under the motivation of previous findings that the optimum
operating point of the 802.11 WLAN lies in nonsaturation
\cite{Zhai05}, we define a simple test for centralized admission
control of multimedia traffic based on queue utilization estimates
of a simple model. Comparison with simulation results for a broad
range of traffic types and load shows that the proposed method
provides an accurate network capacity estimation and the proposed
admission control algorithm prevents both over- and
under-admission problems of previously proposed models.

%% file: cycletime.tex
\section{EDCA Cycle Time Analysis} \label{sec:cycletimeanalysis}

We propose an average cycle time analysis to model the behavior of
the EDCA function of any AC at any station. In this section, we
will first define a Traffic Class (TC). Then, we will derive the
TC-specific average collision probability. Next, we will calculate
the TC-specific average cycle time. Finally, we will relate the
average cycle time and the average collision probability to the
normalized throughput and service time.

The main assumption for saturation analysis is that each AC always
has a frame in service. Note that the performance of EDCA differs
depends on the number of active ACs within the same station as
well as the number of active ACs at the other stations due to the
fact that the EDCA function acts differently in the case of an
internal or an external collision. One of the key differences of
our theoretical formulation from the previous work in the
literature is as follows. We consider both the possibility of a
station running multiple ACs (thus the possibility of internal
collisions) and the possibility of the number of active ACs
varying from station to station. For example, consider a simple
WLAN scenario where an Access Point (AP, labeled $STA_{0}$ in the
sequel) runs 2 downlink ACs, namely AC$_{1}$ and AC$_{2}$.
Similarly, assume $n_{1}$ stations ($STA_{1},\ldots,STA_{n_{1}}$,
$n_{1}>0$) only run AC$_{1}$ and $n_{2}$ other stations
($STA_{n_{1}+1},\ldots,STA_{n_{1}+n_{2}}$, $n_{2}>0$) only have
AC$_{2}$ in the uplink. Although there are 2 distinct ACs active
in the system, the downlink AC$_{i}$ and the uplink AC$_{i}$
($i=\{1,2\}$ for the running example) cannot be expected to have
the same performance due to internal and external collision
differentiation \cite{802.11e}. In this case, the performance
analysis should be carried out individually for 4 different
Traffic Classes (TCs) which are uplink AC$_{1}$, downlink
AC$_{1}$, uplink AC$_{2}$, and downlink AC$_{2}$.

We make the following mathematical definitions for the sake of the
analysis in the sequel.
\begin{itemize}
\item Let $\delta_{k}$ ($0\leq k \leq n_{STA}$) be a vector of
size 4 which denotes the activity status of ACs at $STA_{k}$ where
$n_{STA}$ is the total number of stations that have at least one
active AC. The value at dimension $i$ of $\delta_{k}$ shows
whether AC$_{i}$ is active or not at $STA_{k}$. The entries
corresponding to the indices of active (inactive) ACs are labeled
with 1 (0). In the example above, $\delta_{0} = (0,1,1,0)$,
$\delta_{1} = (0,1,0,0)$, $\delta_{n_{1}+1} = (0,0,1,0)$, etc.
\item Let $\zeta$ be the set of $\delta_{k}$, i.e., $\zeta =
\{\delta_{k}:0\leq k \leq n_{STA}\}$. Above, $\zeta =
\{(0,1,0,0),(0,0,1,0),(0,1,1,0)\}$. \item Let $\psi_{i}$ be the
set of $\delta_{k}$ where AC$_{i}$ is active, i.e.,
$\psi_{i}=\{\delta_{k}:\delta_{k}(i)=1,0\leq k \leq n\}$. In the
example above, $\psi_{1} = \{(0,1,0,0),(0,1,1,0)\}$, $\psi_{2} =
\{(0,0,1,0),(0,1,1,0)\}$, and $\psi_{0}=\psi_{3}=\{\}$.  \item Let
$N(S)$ be an operator on a set $S$ which shows the number of
elements in the set. Then, the total number of TCs with AC$_{i}$
active are $N(\psi_{i})$ and the total number of TCs is
$J=\sum_{i=0}^{i=3}N(\psi_{i})$. Note that $N(\zeta) \leq J$
should always hold. In the sequel, each distinct TC is denoted by
TC$_{j}$ ($0 \leq j < J$). We also define $\sigma_{j}$ as the
activity status vector of TC$_{j}$. In the example above,
$N(\zeta)=3$ and $J=4$. TC$_{0}$ is the AC$_{1}$ when only
AC$_{1}$ is active at the station ($\sigma_{1}=\psi_{1}(1)$).
TC$_{1}$ is the AC$_{1}$ with both AC$_{1}$ and AC$_{2}$ are
active ($\sigma_{2}=\psi_{1}(2)$). TC$_{2}$ is the AC$_{2}$ when
only AC$_{2}$ is active at the station ($\sigma_{3}=\psi_{2}(1)$).
TC$_{3}$ is the AC$_{2}$ with both AC$_{1}$ and AC$_{2}$ are
active ($\sigma_{4}=\psi_{2}(2)$). \item Let $F$ be a function
with the domain of indices of TCs and the range of indices of ACs.
We define this function such as the image of the argument $j$
under function $F$ is the index $i$ of the AC that TC$_{j}$ uses,
i.e., $F(j)=\{i:TC_{j} \in \psi_{i}\}$. \item Let $G$ be a
function from the domain of the indices of TCs to the range of
sets of indices of TCs. We define this function such as the image
of the argument $j$ under mapping $G$ is the set of TC indices
$j'$ with the same $\sigma$, i.e., $G(j)=\{\forall j':
\sigma_{j}=\sigma_{j'}, 0 \leq j' < J\}$.
\end{itemize}

\subsection{TC-specific Average Collision Probability}
\label{sec:collprob}

The difference in AIFS of each AC in EDCA creates the so-called
\textit{contention zones or periods} as shown in
Fig.~\ref{fig:unsat_contzones} \cite{Robinson04},\cite{Hui05}. In
each contention zone, the number of contending TCs may vary. In
order to be consistent with the notation of \cite{802.11e}, we
assume $AIFS_{0}\geq AIFS_{1} \geq AIFS_{2} \geq AIFS_{3}$. Let
$d_{j} = AIFSN_{F(j)} - AIFSN_{3}$. Also, let $n^{th}$ backoff
slot after the completion of $AIFS_{3}$ idle interval following a
transmission period be in contention zone $x$. Then, we define $x
= \max \left( F(y)~|~d_{y} = \underset{z}{\max} (d_{z}~|~d_{z}
\leq n)\right)$ which shows contention zone label $x$ is assigned
the largest index value within a set of ACs that have the largest
AIFSN value which is smaller than or equal to $n+AIFSN_{3}$.

We define $p_{c_{j,x}}$ ($0 \leq j < J$) as the conditional
probability that TC$_{j}$ experiences either an external or an
internal collision in contention zone $x$. Note $AIFS_{x}\geq
AIFS_{F(j)}$ should hold for TC$_{j}$ to transmit in zone $x$.
Following the slot homogeneity assumption of \cite{Bianchi00},
assume that each TC$_{j}$ transmits with constant probability,
$\tau_{j}$. Also, let the total number TC$_{j}$ flows be $f_{j}$.
Then,
\begin{equation}
\label{eq:unsatpcjx} \setlength{\nulldelimiterspace}{0pt}
p_{c_{j,x}} = 1-\frac{\prod \limits_{\forall j':d_{j'}\leq
d_{F^{-1}(x)}} (1-\tau_{j'})^{f_{j'}}}{\prod \limits_{\forall
j'\in G(j):F(j')\leq F(j)}(1-\tau_{j})}.
\end{equation}


We use the Markov chain shown in Fig.~\ref{fig:unsat_AIFSMC} to
find the long term occupancy of the contention zones. Each state
represents the $n^{th}$ backoff slot after the completion of the
AIFS$_{3}$ idle interval following a transmission period. The
Markov chain model uses the fact that a backoff slot is reached if
and only if no transmission occurs in the previous slot. Moreover,
the number of states is limited by the maximum idle time between
two successive transmissions which is
$W_{min}=\min(CW_{F(j),max})$ for a saturated scenario. The
probability that at least one transmission occurs in a backoff
slot in contention zone $x$ is
\begin{equation}
\label{eq:unsatptr} \setlength{\nulldelimiterspace}{0pt}
p^{tr}_{x} = 1-\prod_{\forall j':d_{j'}\leq d_{F^{-1}(x)}}
(1-\tau_{j'})^{f_{j'}}.
\end{equation}

\noindent Note that $F^{-1}$ is not one-to-one. Therefore, we
define the image of $F^{-1}(i)$ as a randomly selected TC index
$j$ which satisfies $F(j)=i$.

Given the state transition probabilities as in
Fig.~\ref{fig:unsat_AIFSMC}, the long term occupancy of the
backoff slots $b'_{n}$ can be obtained from the steady-state
solution of the Markov chain. Then, the TC-specific average
collision probability $p_{c_{j}}$ is found by weighing zone
specific collision probabilities $p_{c_{j,x}}$ according to the
long term occupancy of contention zones (thus backoff slots)
\begin{equation}
\label{eq:unsatpcj}p_{c_{j}} = \frac{\sum_{n=d_{j}+1}^{W_{min}}
p_{c_{j,x}}b'_{n}}{\sum_{n=d_{j}+1}^{W_{min}} b'_{n}}.
\end{equation}
\noindent where $x$ is calculated depending on the value of $n$ as
stated previously.

\subsection{TC-Specific Average Cycle Time} \label{sec:cycletime}

Let $E_{j}[t_{cyc}]$ be average cycle time for a tagged TC$_{j}$
user. $E_{j}[t_{cyc}]$ can be calculated as the sum of average
duration for \textit{i)} the successful transmissions,
$E_{j}[t_{suc}]$, \textit{ii)} the collisions, $E_{j}[t_{col}]$,
and \textit{iii)} the idle slots, $E_{j}[t_{idle}]$ in one cycle.
\begin{equation} \label{eq:cycletime}
E_{j}[t_{cyc}] = E_{j}[t_{suc}] + E_{j}[t_{col}] + E_{j}[t_{idle}]
\end{equation}

In order to calculate the average time spent on successful
transmissions during a TC$_{j}$ cycle time, we should find the
expected number of total successful transmissions between two
successful transmissions of TC$_{j}$. Let $Q_{j}$ represent this
random variable. Also, let $\gamma_{j}$ be the probability that
the transmitted packet belongs to an arbitrary user from TC$_{j}$
given that the transmission is successful. Then,
\begin{equation}\label{eq:gamma_j}
\gamma_{j} = \frac{\sum_{n=d_{j}+1}^{W_{min}}
b'_{n}p_{s_{j,n}}/f_{j}}{\sum_{n=d_{j}+1}^{W_{min}}
(b'_{n}\sum\limits_{\forall l} p_{s_{l,n}})}
\end{equation}
\noindent where
\begin{equation}\label{eq:p_s_j_cycle}
p_{s_{j,n}} =
\left\{ \\
\begin{IEEEeqnarraybox}[\relax][c]{lc}
\frac{f_{j}\tau_{j}\prod_{j':d_{j'}\leq
n-1}(1-\tau_{j'})^{f_{j'}}}{\prod_{\forall j'\in
G(j):F(j')\leq F(j)}(1-\tau_{j})}, &~{\rm if}~n \geq d_{j}+1 \\
0, &~{\rm if }~n < d_{j}+1.
\end{IEEEeqnarraybox}
\right.
\end{equation}


Then, the Probability Mass Function (PMF) of $Q_{j}$ is
\begin{equation}\label{eq:PMFsucctrans}
Pr(Q_{j}=k) = \gamma_{j}(1-\gamma_{j})^{k}, ~~k \geq 0.
\end{equation}

We can calculate the expected number of successful transmissions
of any TC$_{j'}$ during the cycle time of TC$_{j}$, $ST_{j',j}$,
as
\begin{equation}\label{eq:ExpectedindividualAC}
ST_{j',j} = f_{j'}E[Q_{j}] \frac{\gamma_{j'}}{1-\gamma_{j}}.
\end{equation}

Inserting $E[Q_{j}]=(1-\gamma_{j})/\gamma_{j}$ in
(\ref{eq:ExpectedindividualAC}), the intuition that each user from
TC$_{j}$ can transmit successfully once on average during the
cycle time of another TC$_{j}$ user, i.e., $ST_{j,j}=f_{j}$, is
confirmed.
Including the own successful packet
transmission time of tagged TC$_{j}$ user in $E_{j}[t_{suc}]$, we
find
\begin{equation}\label{eq:Etsuc}
E_{j}[t_{suc}] = \sum_{\forall j'} ST_{j',j}T_{s_{j'}}
\end{equation}

\noindent where $T_{s_{j'}}$ is defined as the time required for a
successful packet exchange sequence (will be derived in
(\ref{eq:unsatTs})).

To obtain $E_{j}[t_{col}]$, we need to calculate the average
number of users who are involved in a collision, $f_{c_{n}}$, at
the $n^{th}$ slot after last busy time for given $f_{j}$ and
$\tau_{j}$, $\forall j$. Let the total number of users
transmitting at the $n^{th}$ slot after last busy time be denoted
as $Y_{n}$. We see that $Y_{n}$ is the sum of random variables,
$Binomial(f_{j},\tau_{j})$, $\forall j:~d_{j}\leq n-1$. Employing
simple probability theory, we can calculate
$f_{c_{n}}=E[Y_{n}|Y_{n}\geq 2]$. After some algebra and
simplification,
\begin{equation}
f_{c_{n}} = \frac{\sum\limits_{j:d_{j}\leq n-1}
(f_{j}\tau_{j}-p_{s_{j,n}})}{1-\prod\limits_{j:d_{j}\leq
n-1}(1-\tau_{j})^{f_{j}}-\sum\limits_{j:d_{j}\leq n-1}p_{s_{j,n}}}
\end{equation}

If we let the average number of users involved in a collision at
an arbitrary backoff slot be $f_{c}$, then
\begin{equation}
f_{c} = \sum_{\forall n} b'_{n}f_{c_{n}}.
\end{equation}

We can also calculate the expected number of collisions that an
TC$_{j'}$ user experiences during the cycle time of a TC$_{j}$,
$CT_{j',j}$, as
\begin{equation} \label{eq:CTjj}
CT_{j',j} = \frac{p_{c_{j'}}}{1-p_{c_{j'}}}ST_{j',j}.
\end{equation}

\noindent Then, defining $T_{c_{j'}}$ as the time wasted in a
collision period (will be derived in (\ref{eq:unsatTc})),
\begin{equation} \label{eq:Ejtcol}
E_{j}[t_{col}] = \frac{1}{f_{c}}\sum_{\forall j'}
CT_{j',j}T_{c_{j'}}.
\end{equation}

Given $p_{c_{j}}$, we can calculate the expected number of backoff
slots $E_{j}[t_{bo}]$ that TC$_{j}$ waits before attempting a
transmission. Let $W_{i,k}$ be the CW size of AC$_{i}$ at backoff
stage $k$ \cite{Inan07_ICC}. Note that, when the retry limit
$r_{i}$ is reached, any packet is discarded. Therefore, another
$E_{j}[t_{bo}]$ passes between two transmissions with probability
$p_{c_{j}}^{r_{i}}$ (where $i=F(j)$).
\begin{equation}\label{eq:aveBO}
E_{j}[t_{bo}]=\frac{1}{1-p_{c_{j}}^{r_{i}}}\sum_{k=1}^{r_{i}}p_{c_{j}}^{k-1}(1-p_{c_{j}})\frac{W_{i,k}}{2}.
\end{equation}

\noindent Noticing that between two successful transmissions,
AC$_{j}$ also experiences $CT_{j,j}$ collisions,
\begin{equation}\label{eq:E_j_t_idle}
E_{j}[t_{idle}] = E_{j}[t_{bo}](CT_{j,j}/f_{j}+1)t_{slot}.
\end{equation}

The transmission probability of a user using TC$_{j}$ is
\begin{equation}\label{eq:tauapp}
\tau_{j} = \frac{1}{E_{j}[t_{bo}]+1}.
\end{equation}

\noindent Note that, in \cite{Hui05}, it is proven that the mean
value analysis for the average transmission probability calculated
as in (\ref{eq:tauapp}) matches the Markov analysis of
\cite{Bianchi00}.

The equations (\ref{eq:unsatpcjx})-(\ref{eq:unsatpcj}),
(\ref{eq:aveBO}), and (\ref{eq:tauapp}) are a set of nonlinear
equations which can be solved numerically for $\tau_{j}$ and
$p_{c_{j}}$, $\forall j$. Then, the average cycle time for
AC$_{j}$, $\forall j$, can be calculated using
(\ref{eq:cycletime}) where each term in (\ref{eq:cycletime}) is
obtained via (\ref{eq:gamma_j})-(\ref{eq:E_j_t_idle}).

\subsection{Performance Analysis} \label{sec:performance}

Let $T_{p_{j}}$ be the average payload transmission time for
TC$_{j}$ ($T_{p_{j}}$ includes the transmission time of MAC and
PHY headers), $\delta$ be the propagation delay, $T_{ack}$ be the
time required for acknowledgment packet (ACK) transmission. Then,
for the basic access scheme, we define the time spent in a
successful transmission $T_{s_{j}}$ and a collision $T_{c_{j}}$
for any TC$_{j}$ as
\begin{align}\label{eq:unsatTs}
T_{s_{j}} = & T_{p_{j}} + \delta + SIFS + T_{ack} + \delta +
AIFS_{F(j)}
\\ \label{eq:unsatTc} T_{c_{j}} = & T_{p^{*}_{j}} + ACK\_Timeout +
AIFS_{F(j)}
\end{align}
\noindent where $T_{p^{*}_{j}}$ is the average transmission time
of the longest packet payload involved in a collision
\cite{Bianchi00}. For simplicity, we assume the packet size to be
equal for any TC, then $T_{p^{*}_{j}}=T_{p_{j}}$. Being not
explicitly specified in the standards, we set $ACK\_Timeout$,
using Extended Inter Frame Space (EIFS) as $EIFS_{i}-AIFS_{i}$
($i=F(j)$). Note that the extensions of~(\ref{eq:unsatTs})
and~(\ref{eq:unsatTc}) for the RTS/CTS scheme are straightforward
\cite{Bianchi00}.

The average cycle time of an TC represents the renewal cycle for
each TC. Then, the normalized throughput of TC$_{j}$ is defined as
the successfully transmitted information per renewal cycle
\begin{equation}\label{eq:Sj_cycle}
S_{j} = \frac{f_{j}T_{p_{j}}}{E_{j}[t_{cyc}]}.
\end{equation}

The TC-specific cycle time is directly related but not equal to
the mean protocol service time. By definition, the cycle time is
the duration between successful transmissions. We define the
average protocol service time such that it also considers the
service time of packets which are dropped due to retry limit. On
the average, $1/p_{j,drop}$ service intervals correspond to
$1/p_{j,drop}-1$ cycles (where $p_{j,drop} = p_{c_{j}}^{r_{i}}$ is
the average packet drop probability). Therefore, the mean service
time $E_{j}[t_{srv}]$ can be calculated as
\begin{align}\label{eq:pdp_cycle}
E_{j}[t_{srv}] = (1-p_{j,drop})E_{j}[t_{cyc}].
\end{align}


%% file: validation.tex
\subsection{Validation} \label{sec:modelvalidation}

We validate the accuracy of the numerical results by comparing
them to the simulation results obtained from ns-2 \cite{ns2}. For
the simulations, we employ the IEEE 802.11e HCF MAC simulation
model for ns-2.28 \cite{ourcode}. This module implements all the
EDCA and HCCA functionalities stated in \cite{802.11e}.

In simulations, we consider two ACs, one high priority (AC$_{3}$)
and one low priority (AC$_{1}$). Unless otherwise is stated, each
station runs only one AC. For both ACs, the payload size is 1000
bytes. RTS/CTS handshake is turned on. The simulation results are
reported for the wireless channel which is assumed to be not prone
to any errors during transmission. The errored channel case is
left for future study. All the stations have 802.11g Physical
Layer (PHY) using 54 Mbps and 6 Mbps as the data and basic rate
respectively ($T_{slot}=9~\mu s$, $SIFS=10~\mu s$) \cite{802.11g}.
The simulation runtime is 100 seconds.

In the first set of experiments, we set $AIFSN_{1}=3$,
$AIFSN_{3}=2$, $CW_{1,min}=31$, $CW_{3,min}=15$, $m_{1}=m_{3}=3$,
$r_{1}=r_{3}=7$. Fig.~\ref{fig:A1_thp_GC07} shows the normalized
throughput of each AC when both $N_{1}$ and $N_{3}$ are varied
from 5 to 30 and equal to each other. As the comparison with a
more detailed analytical model \cite{Inan07_ICC} and the
simulation results reveal, the cycle time analysis can predict
saturation throughput accurately. Fig.~\ref{fig:A1_mpst_GC07}
displays the mean protocol service time for the same scenario of
Fig.~\ref{fig:A1_thp_GC07}. As comparison with \cite{Inan07_ICC}
and the simulation results show, the mean protocol service time
can accurately be predicted by the proposed cycle time model.
Although not included in the figures, a similar discussion holds
for the comparison with other detailed and/or complex models of
\cite{Tao06}-\cite{Banchs06} and for other performance metrics
such as mean packet drop probability.

In the second set of experiments, we fix the EDCA parameters of
one AC and vary the parameters of the other AC in order to show
the proposed cycle time model accurately captures the normalized
throughput for different sets of EDCA parameters. In the
simulations, both $N_{1}$ and $N_{3}$ are set to 10.
Fig.~\ref{fig:v_aifs_cw_1_thp_GC07} shows the normalized
throughput of each AC when we set $AIFSN_{3}=2$, $CW_{3,min}=15$,
and vary $AIFSN_{1}$ and $CW_{1,min}$.
Fig.~\ref{fig:v_aifs_cw_3_thp_GC07} shows the normalized
throughput of each AC when we set $AIFSN_{1}=4$, $CW_{1,min}=127$,
and vary $AIFSN_{3}$ and $CW_{3,min}$. As the comparison with
simulation results show, the predictions of the proposed cycle
time model are accurate. Although, we do not include the results
for packet drop probability and service time for this experiment,
no discernable trends toward error are observed.

In the third set of experiments, we test the performance of the
system when the stations run multiple ACs so that virtual
collisions may occur. The stations run only AC$_{1}$, only
AC$_{3}$, or both. Like in Section \ref{sec:cycletimeanalysis}, we
define TC$_{0}$ as the AC$_{3}$ when only AC$_{3}$ is active at
the station, TC$_{1}$ the AC$_{3}$ when both AC$_{3}$ and AC$_{1}$
are active at the station, TC$_{2}$ as the AC$_{1}$ when only
AC$_{1}$ is active at the station, and TC$_{3}$ the AC$_{1}$ when
both AC$_{3}$ and AC$_{1}$ are active at the station. We keep both
$N_{1}$ and $N_{3}$ at 10, and vary the number of TC$_{1}$ and
TC$_{3}$ from 0 to 10 (therefore, TC$_{0}$ and TC$_{2}$ vary from
10 to 0). Fig.~\ref{fig:multipleAC_TC} shows the normalized
throughput of each TC. The predictions of the proposed analytical
model follow the simulation results closely. Although not
significant for the tested scenario and not apparent in the
graphical results, a closer look on the numerical results present
the (slightly) higher level of differentiation between AC$_{3}$
and AC$_{1}$ which is due to the additional prioritization
introduced at the virtual collision procedure.

%% file: admission.tex
\section{Multimedia Capacity Analysis for 802.11e Infrastructure BSS}\label{sec:cacEDCA}

When working in the saturated case, the contention-based 802.11
MAC suffers from a large collision probability, which leads to low
channel utilization and excessively long delay. As shown in
\cite{Zhai05}, the optimal operating point for the 802.11 to work
lies in nonsaturation where contention-based 802.11 MAC can
achieve maximum throughput and small delay. In \cite{Serrano06},
it is also shown that a very small increase in system load yields
a huge increase (of about two orders of magnitude) of the backoff
delay. When the traffic load does not exceed the service rate at
saturation, the resulting medium access delay is very small.

In this section, we propose a novel framework where we calculate
TC-specific average frame service rate $\mu$ via a weighted
summation of saturation service rate $E[t_{srv}]$ over varying
number of active stations. Defining a TC-specific average queue
utilization ratio $\rho$, we design a simple call admission
control algorithm which limits the number of admitted real-time
multimedia flows in the 802.11e infrastructure BSS in order to
prevent the corresponding TC queues going into saturation. As
specified in \cite{802.11e}, the admission control is conducted at
the AP. Admitted real-time multimedia flows can be served with QoS
guarantees, since low transmission delay and packet loss rate can
be maintained when the 802.11e WLAN is in nonsaturation
\cite{Zhai05},\cite{Serrano06}. Comparing with simulation results,
we show that not only does the proposed admission control
algorithm prevent the so-called over-admission or under-admission
problems but also efficiently utilizes the network capacity.

\subsection{TC-specific Average Queue Utilization
Ratio}\label{sec:queueutilization}

Each station runs a QoS reservation procedure with the AP for all
of its traffic streams that need parameterized (guaranteed) QoS
support. The Station Management Entity (SME) at the AP decides
whether the Traffic Stream (TS) is admitted or not regarding the
Traffic Specification (TSPEC) in the Add Traffic Stream (ADDTS)
request provided by the station. The TSPEC specifies the Traffic
Stream Identification Number ($TSID$), the user priority ($UP$),
the mean data rate ($R$), and the mean packet size ($L$) of the
corresponding TS \cite{802.11e}.

Let average frame service rate for TC$_{j}$ be denoted as
$\mu_{j}$. Also let the average packet arrival rate for TC$_{j}$
be denoted as $\lambda_{j}$ which can easily be calculated
employing $R$ of TSs using the same TC at the same station. For
simplicity, in the sequel, we assume that TCs at different
stations are running TSs with equal TSPEC values (so all traffic
parameters remain TC-specific). Though all work in this section
can be generalized for varying traffic load and parameters within
a TC vary at different stations, we opted not to present this
out-of-scope generalization since it would make the model
difficult to understand.

We define TC-specific queue utilization ratio $\rho$ as follows
\begin{equation} \label{eq:rho}
\rho_{j} = \lambda_{j} / \mu_{j},~ \forall j.
\end{equation}

\subsection{TC-specific Average Frame Service
Rate}\label{sec:servicetime}

The TC-specific average queue utilization ratio $\rho_{j}$ shows
the percentage of time on average that TC$_{j}$ has a frame in
service. In other words, $\rho_{j}$ is the probability that
TC$_{j}$ is active. Our novel approach in calculating $\mu_{j}$ is
forming a weighted summation of $E_{j}[t_{srv}]$ for varying
number of active TCs.

Let
$P_{TC_{0},TC_{1},...,TC_{j},...,TC_{J-1}}^{j}(f'_{0},f'_{1},...,f'_{j},...,f'_{J-1})$
denote the joint conditional probability that $f'_{j}$ stations
using TC$_{j}$, $\forall j$, are active given that one TC$_{j}$
has a frame in service and the total number of TCs is $J$. Also,
let $E_{j}[t_{srv}(f'_{0},f'_{1},...,f'_{J-1})]$ denote the
average service time when $f'_{j}$ stations using TC$_{j}$,
$\forall j$, are active. We use the proposed cycle time model in
Section \ref{sec:cycletimeanalysis} to calculate
$E_{j}[t_{srv}(f'_{0},f'_{1},...,f'_{J-1})]$\footnotemark\footnotetext{The
proposed capacity estimation framework is generic. Any other
accurate saturation analysis method can also be employed for
calculating the service time.}. Then, the TC-specific average
frame service rate $\mu_{j}$ is calculated as follows
\begin{equation}\label{eq:mu}
\frac{1}{\mu_{j}} =\sum_{0 \leq f'_{0}\leq f_{0}}...\sum_{1 \leq
f'_{j} \leq f_{j}}...\sum_{0\leq f'_{J-1}\leq f_{J-1}}
E_{j}[t_{srv}(f'_{0},...,f'_{j},...,f'_{J-1})]\cdot
P_{TC_{0},...,TC_{j},...,TC_{J-1}}^{j}(f'_{0},...,f'_{j},...,f'_{J-1}).
\end{equation}

Note that the case when $\sum_{j'=0}^{J-1}f'_{j'}=1$, i.e., there
is only one active TC, is not considered by the proposed cycle
time model. On the other hand, the cycle time calculation in this
case is straightforward. Since no collisions can occur and no
other station is active, the successful transmission is performed
at AIFS completion. Therefore,
$E_{j}[t_{srv}(f'_{0},f'_{1},...,f'_{J-1})]=T_{s_{j}}$ if
$\sum_{j'=0}^{J-1}f'_{j'}=1$.

We noticed that the distribution of the number of active TCs
approximates the sum of independent Binomial distributions with
parameters $f_{j'}$ and $\rho_{j'}$, $\forall j'$ as in
(\ref{eq:activitydist}) for the traffic models we used in this
study. We confirm the validity of (\ref{eq:activitydist}) via
comparing the analytical estimations with simulation results in
Section \ref{sec:cacvalidation}. On the other hand, we do not
argue that the binomial activity distribution holds for any type
of traffic model in any scenario. Our observation is that for
widely used voice and video traffic models this approximation
works well. The proposed framework is generic in the sense that
any other activity distribution profile may be used to incorporate
other traffic models in other network scenarios.

\begin{equation}\label{eq:activitydist}
P_{TC_{0},TC_{1},...,TC_{j},...,TC_{J-1}}^{j}(f'_{0},f'_{1},...,f'_{j},...,f'_{J-1})
=
\dbinom{f_{j}-1}{f'_{j}-1}\rho_{j}^{f'_{j}-1}(1-\rho_{j})^{f_{j}-f'_{j}}\prod_{\forall
j':j'\neq j}
\dbinom{f_{j'}}{f'_{j'}}\rho_{j'}^{f'_{j'}}(1-\rho_{j'})^{f_{j'}-f'_{j'}}
\end{equation}

The fixed-point equations (\ref{eq:rho})-(\ref{eq:activitydist})
can numerically be solved for $\rho_{j}$ and $\mu_{j}$, $\forall
j$.

\subsection{Admission Control Procedure}\label{sec:admissionproc}

Upon receiving the ADDTS request, the AP associates the TS with
the AC and the TC using the value in the $UP$ field and the
station MAC address. The traffic stream is admitted if and only if
the following tests succeed
\begin{equation} \label{eq:admissiontest}
\rho_{j} \leq \rho_{th}, ~ \forall j.
\end{equation}

\noindent where $\rho_{th} \leq 1$. The tests in
(\ref{eq:admissiontest}) ensure that the average traffic arrival
rate to all TCs is smaller than the average service rate that can
be provided to them. Therefore, the MAC queues of all TCs can be
considered to be stable (all TCs remain in nonsaturation on
average).

When a real-time flow ends, the source node transmits a Delete
Traffic Stream (DELTS) request for the TS \cite{802.11e}. The AP
deletes the corresponding entry from the list of admitted flows.

A few remarks on admission control and capacity analysis are as
follows.
\begin{itemize}
\item The proposed capacity analysis and admission control scheme can
easily be extended to the case where some TCs are running
best-effort traffic. We actually do a worst-case analysis in
Section \ref{sec:cacvalidation} where the TCs that run best-effort
traffic are assumed to be always active. This generalizes
(\ref{eq:mu}) as
\begin{equation}\label{eq:mu_general}
\frac{1}{\mu_{j}} =\sum_{0 \leq f'_{1}\leq f_{1}}...\sum_{1 \leq
f'_{j}\leq f_{j}}...\sum_{0\leq f'_{J'}\leq f_{J'}}
E_{j}[t_{srv}(f'_{1},...,f'_{j},...,f'_{J'-1},f_{J'},...,f_{J-1})]\cdot
P_{TC_{1},...,TC_{J'}}^{j}(f'_{1},...,f'_{J'})
\end{equation}
\noindent where $J'$ and $J-J'$ are the number of TCs that run
multimedia and best-effort flows respectively. In this case, the
admission control tests in (\ref{eq:admissiontest}) are done for
TCs that run real-time flows, i.e., $0 \leq j \leq J'$.

\item Although the employed saturation model does not consider wireless channel
errors, the admission control scheme can still be effective in an
error-prone wireless channel as the admission control decisions
are threshold-based. Selecting a comparably smaller $\rho_{th}<1$
can provide the necessary room for packet retransmissions occuring
as a result of wireless channel losses. This may be a more simpler
approach when compared to a solution that includes the design of a
more complex saturation analysis model considering wireless
channel errors. The investigation is left as future work.

\item The proposed capacity estimation scheme is solely based on mean values and do not consider the
worst-case scenario where all the admitted Variable Bit Rate (VBR)
multimedia traffic may instantaneously transmit at their peak rate
($R_{peak}$). Again a wise decision $\rho_{th}$ can limit the
channel utilization by multimedia flows thereby leaving room to
accommodate bandwidth fluctuations caused by VBR traffic.
Alternatively, $R_{peak}$ may be used in the calculation of
$\lambda$ in (\ref{eq:rho}). On the other hand, when $R_{peak}/R$
is very large, this may result in the rejection of many multimedia
flows and unnecessarily low channel utilization.
\item The TSPECs may also specify a Delay
Bound ($DB$) which denotes the maximum time allowed to transport
the frames across the wireless interface including the queueing
delay \cite{802.11e}. As also provided in \cite[Table I]{Zhai05},
multimedia services should satisfy QoS requirements in terms of
one-way transmission delay, delay variation, and packet loss rate.
For example, for voice and video the excellent (acceptable)
quality is satisfied if the delay is smaller than 150 ms (400 ms)
and the packet loss rate is smaller than $1\%-3\%$ \cite{Zhai05}.
Note that packet loss rate includes the dropped packets at the
playout buffer of the receiver when the packets are not received
within the delay bound. Our capacity analysis does not explicitly
consider these metrics in admission control. On the other hand,
the proposed call admission control algorithm makes the multimedia
TC queues remain stable (TC queues do not go into saturation) by
limiting the number of admitted real-time flows. This provides low
transmission delays and packet loss ratio due to the low collision
probability in nonsaturation \cite{Zhai05}.
\item In the simulations, we
observed that the delay experienced by multimedia flows in
nonsaturation can go up to 40-50 ms depending on the scenario. In
order to guarantee a stochastic delay bound, the admission control
tests in (\ref{eq:admissiontest}) should be extended. We may use
the method proposed in \cite{Cheng06}
\begin{equation}
{\rm Pr}(Q_{j}>DB_{j}\cdot \mu_{j}) \leq\epsilon
\end{equation}
\noindent where $Q_{j}$ is the queue length of the TC$_{j}$ and
$\epsilon$ is the delay violation probability. This test can be
extended further for on/off traffic sources and statistical
multiplexing at the AP as shown in \cite{Cheng06}. On the other
hand, in the simulation scenarios we have studied, the addition of
this test does not limit the already admitted traffic using
(\ref{eq:admissiontest}) since the QoS requirements of the
multimedia flows are always satisfied if the system is in
nonsaturation state. \item In the simulations, we consider two
types of traffic sources; voice and video, where the average
packet size of different traffic sources vary. Therefore, $T_{c}$
is not equal for any TC since $T_{p_{j}^{*}}=T_{p_{j}}$ does not
always hold. Due to space limitations, we do not include the
calculation of $T_{c_{j}}$ in this case. We use the method in
\cite{Bianchi00} which has an extensive treatment of the subject.
\end{itemize}

\subsection{Validation}\label{sec:cacvalidation}


For the experiments, we use a network topology such that any
connection is initiated between a distinct party in the Internet
and the WLAN. The traffic is relayed at the AP from (to) the
wireless channel to (from) the wired link. The simulations
consider three types of traffic sources; voice, video, and
background data. The voice traffic models G.711 or G.729 VoIP
application as Constant Bit Rate (CBR) traffic (without the use of
silence suppression scheme). The CBR traffic model is used for two
reasons; \textit{i)} it provides a worst-case upper bound for the
case when the traffic presents on-off traffic characteristics
(silence suppression) and \textit{ii)} this enables comparison of
voice capacity results with the models proposed in
\cite{Hui06,Medepalli05_2,Hole04,Garg03_1,Garg03_2,Cai06,Gao06}.
The parameters of the VoIP codecs are set as in \cite[Table
I]{Cai06}. For the video source models, we use the traces of real
MPEG-4 video streams \cite{Seeling04}. For the particular video
source used in the simulations presented in this paper, the
average codec bit rate is 174 kbps with an average packet size of
821 bytes. Real-time packets have 40-byte length RTP/UDP/IP
header. The background data traffic is modeled by bulk data
transfer where every AC using this type of traffic is saturated.
Voice flows use AC$_{3}$, video flows use AC$_{2}$, and background
traffic uses AC$_{1}$. We set the EDCA parameters as suggested in
\cite{802.11e}; $AIFSN_{3}=3$, $AIFSN_{2}=2$, $AIFSN_{3}=2$,
$CW_{1,min}=31$, $CW_{2,min}=15$, $CW_{3,min}=7$,
$CW_{1,max}=1023$, $CW_{2,max}=31$, $CW_{3,max}=15$, $r=7$. PHY
parameters are set as stated in Section \ref{sec:modelvalidation}.
The wired link delay is set to 20 ms for all connections.


\subsubsection{Voice Capacity Analysis}

In the first set of experiments, we investigate the VoIP capacity
of 802.11e WLAN when no other type of traffic coexists. A two-way
voice connection is established every $\omega$ ms, with the
starting time randomly chosen over $[0,\omega]$ ms. We set
$\omega$ equal to the packet interval duration of the voice codec
used. Table \ref{tab:VoIPcapacity} tabulates the maximum number of
admitted VoIP connections for different codecs. In the
simulations, the maximum number of voice connections is obtained
in such a way that one more connection results in a packet loss
ratio\footnotemark\footnotetext{A packet drop occurs at the source
if the packet cannot be delivered successfully in the maximum
limit of retries, $r$, or there is no available room for the
packet in the MAC buffer, and at the sink if the end-to-end delay
for the delivered packet exceeds 150 ms \cite{Cai06}.} larger than
$1\%$. As shown in Table \ref{tab:VoIPcapacity}, the analytical
results for the proposed model and the simulation results closely
follow each other. As the comparison of the performance with
802.11e MAC parameters in 802.11g PHY layer indicates, the model
in \cite{Cai06} has significant under-admission problems for an
arbitrary selection of MAC parameters (especially when the CW
settings are small and the underlying PHY is
802.11g)\footnotemark\footnotetext{Our implementation of the model
in \cite{Cai06} duplicates the analytical results in \cite{Cai06}
which are for a specific DCF MAC parameter set. Although the
results are not provided in this paper, the analytical results for
the proposed model and our simulation results also confirm the
capacity prediction of \cite{Cai06} for the specific DCF
scenario.}.
We do not provide any comparison with
\cite{Hole04,Garg03_1,Garg03_2} the over-admission problems of
which are already shown in \cite{Cai06}.

Fig.~\ref{fig:simG711_plr_delay} shows the packet loss ratio and
the average delay of successfully delivered
packets\footnotemark\footnotetext{The presented average delay
results are only for the wireless link and excludes the wired link
delay. The delay for packets that are not delivered within an
end-to-end delay (sum of wireless and wired link delays) of 150 ms
are not included in the average delay calculation.} for increasing
number of active G.711 VoIP connections and codec packet sample
interval. These results are obtained via simulation. As the
comparison of the results in Table~\ref{tab:VoIPcapacity} and
Fig.~\ref{fig:simG711_plr_delay} denotes, there is a sudden
increase in the downlink packet loss ratio and the average
downlink packet delay mainly due to the increasing queueing delay
when the queue utilization ratio exceeds the threshold,
$\rho_{th}$=1\footnotemark\footnotetext{In the simulations, the
MAC buffer size for each node is set to 100 packets. The packet
loss ratio and the average delay for successfully delivered
packets depend on the buffer size. When the buffer size is
smaller, the packet loss ratio is larger and the average delay for
successfully delivered packets is smaller. As we have confirmed
via simulations (specifically, when the buffer size is 20
packets), the capacity in terms of number of flows stays the
same.}. When the load does not exceed the capacity, the packet
loss ratio stays smaller than $1\%$ and the average wireless link
delay is around 10 ms. In the experiments, the downlink always
suffer longer queueing delays and is the main limitation on VoIP
capacity. The uplink experiences comparably much smaller packet
delays and much less packet losses. We do not include uplink
results in Fig.~\ref{fig:simG711_plr_delay} in order not to crowd
the figure. Although the corresponding results are not presented,
a similar discussion holds when VoIP flows employ G.729 codec.

Fig.~\ref{fig:activity_pdf} shows the probability density function
(pdf) of active number of TCs given that the TC at the AP or at
the non-AP station (denoted as STA in the figure) is active in a
scenario consisting of only VoIP connections (G.711 VoIP codec
with 10 ms packet intervals). As previously stated, we
analytically calculate
$P_{TC_{0},TC_{1},...,TC_{j},...,TC_{J-1}}^{j}(f'_{0},f'_{1},...,f'_{j},...,f'_{J-1})$
by assuming that the distribution of the number of active TCs
approximates a Binomial distribution with parameters $f_{j}$ and
$\rho_{j}$. The comparison in Fig.~\ref{fig:activity_pdf} shows
that the pdf of analytical calculation closely follows the pdf
obtained through simulation. Although the results are not
presented here, a similar discussion holds for other codecs with
different packet interval values. The pdf results for simulations
are obtained through averaging over several simulation runs with
different random number generator seeds and randomized flow start
times.

\subsubsection{Voice Capacity Analysis in the Presence of Background
Traffic} In the second set of experiments, we investigate the VoIP
capacity when heavy background traffic coexists. Note that the
analytical models of
\cite{Medepalli05_2,Hole04,Garg03_1,Garg03_2,Cai06,Gao06,Cheng06}
do not provide such analysis capability. Table
\ref{tab:VoIPdatacapacity} shows the number of admitted G.711 VoIP
flows for increasing the number of two-way background data
connections. The comparison of analytical and simulation results
shows that the proposed admission control scheme is highly
accurate when a number of TCs (background) are always assumed
active while some others (VoIP) are in nonsaturation. As the
comparison of Tables \ref{tab:VoIPcapacity} and
\ref{tab:VoIPdatacapacity} presents, the coexistence of background
traffic is a big hit on the multimedia capacity of the WLAN. When
the number of data connections is 5, the number of admitted flows
decreases by around $30\%$. The decrease ratio goes up to $60\%$
when the number of data connections is increased to 30.
Interestingly, the decrease ratio is almost insensitive to packet
sampling interval length. Although the results are not presented,
a similar discussion holds when VoIP flows employ G.729 codec.

\subsubsection{Voice and Video Capacity Analysis}
In the third set of experiments, we investigate the capacity of
802.11e WLAN when both voice and video traffic coexist (using
different ACs). Once again, note that the analytical models of
\cite{Medepalli05_2,Hole04,Garg03_1,Garg03_2,Cai06,Gao06,Cheng06}
do not provide such analysis capability. Table
\ref{tab:VoIPvideocapacity} shows the number of admitted uplink,
downlink, and two-way MPEG-4 flows for increasing the number of
VoIP connections. In this scenario, we use the G.711 codec with a
20 ms sample interval. As the results indicate, the analytical and
simulation results closely follow each other. Such a comparison
reveals that the proposed capacity prediction and admission
control scheme is also effective when different classes of
multimedia traffic coexist in the BSS. As the comparison of the
number of admitted uplink and downlink flows shows, channel
contention overhead in the main limitation on capacity. For the
same number of coexisting VoIP connections, the number of admitted
downlink flows is larger than the number of admitted uplink flows,
as contention overhead is much lower in the downlink scenario.
With increasing number of VoIP connections, the difference
increases as well. As expected, the two-way video capacity in
terms of admitted number of flows is less than the capacity in the
uplink only and the downlink only scenarios. The increasing VoIP
load does not affect the ratio of downlink to two-way video
capacity as significantly as it affects the ratio of the ratio of
downlink to uplink video capacity.

%% file: conclusion.tex
\section{Conclusion}

We have developed a simple and novel average cycle time model to
evaluate the performance of the EDCA function in saturation. The
proposed model captures the performance in the case of an
arbitrary assignment of AC-specific AIFS and CW values and is the
first model to consider an arbitrary distribution of active ACs at
the stations. We have shown that the analytical results obtained
using the cycle time model closely follow the accurate predictions
of the previously proposed more complex analytical models and
simulation results. The proposed cycle time analysis can serve as
a simple and practical alternative model for EDCA saturation
throughput analysis.

We have also designed a practical and simple multimedia capacity
prediction and admission control algorithm to limit the number of
admitted realtime multimedia flows in the 802.11e infrastructure
BSS. Motivated by the previous findings in the literature such
that the contention-based 802.11 MAC can achieve high throughput
and low delay in nonsaturation, the proposed admission control
algorithm is based on simple tests on station- and AC-specific
queue utilization ratio estimates. Our novel approach is the
calculation of the queue utilization ratio by weighing the average
service time predictions of the proposed cycle time saturation
model among varying number of active stations. The proposed simple
framework is effective in capacity estimation in the case of
coexisting multimedia flows using different ACs with arbitrarily
selected MAC parameters. Comparing the theoretical results with
simulations, we have shown that the proposed algorithm provides
guaranteed QoS for coexisting voice or video connections. One of
the key insights provided by this study is the accuracy of the
proposed approximate capacity estimation framework that uses
relatively simpler saturation analysis rather than defining a more
complex and hard to implement nonsaturation model.

%% file: figures.tex
\clearpage
\begin{figure}
\center{\epsfig{file=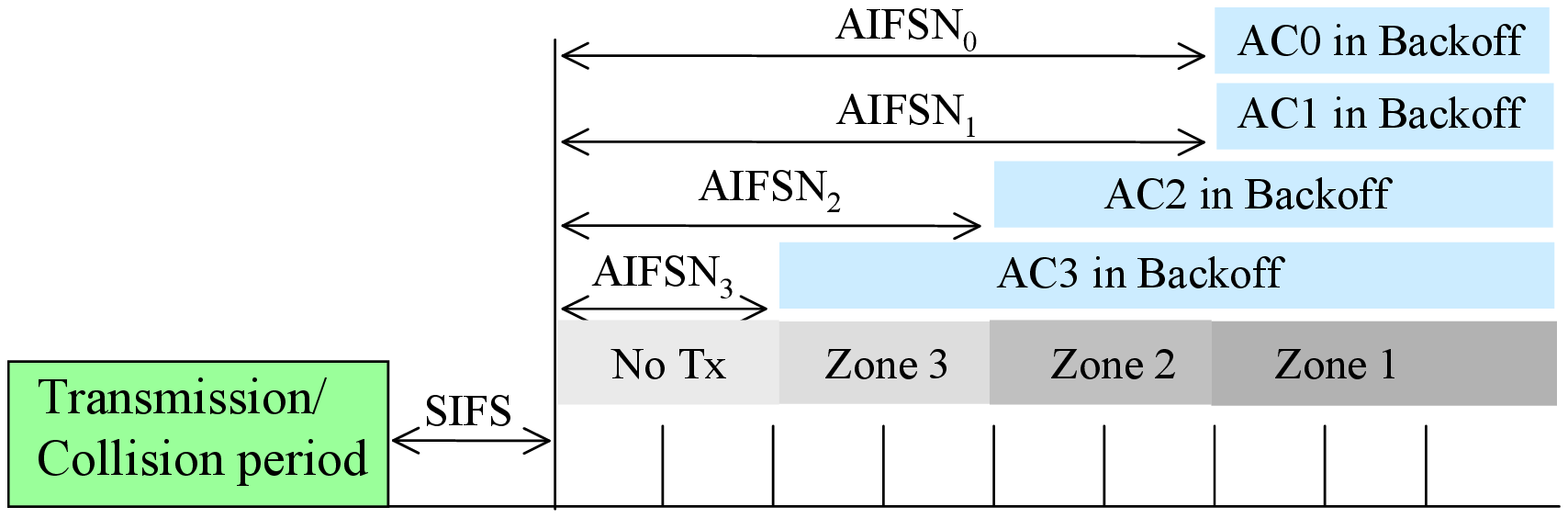,height=17
cm,angle=-90}} \caption[] {\label{fig:unsat_contzones} EDCA
backoff after busy medium. }
\end{figure}

\clearpage
\begin{figure}
\center{\epsfig{file=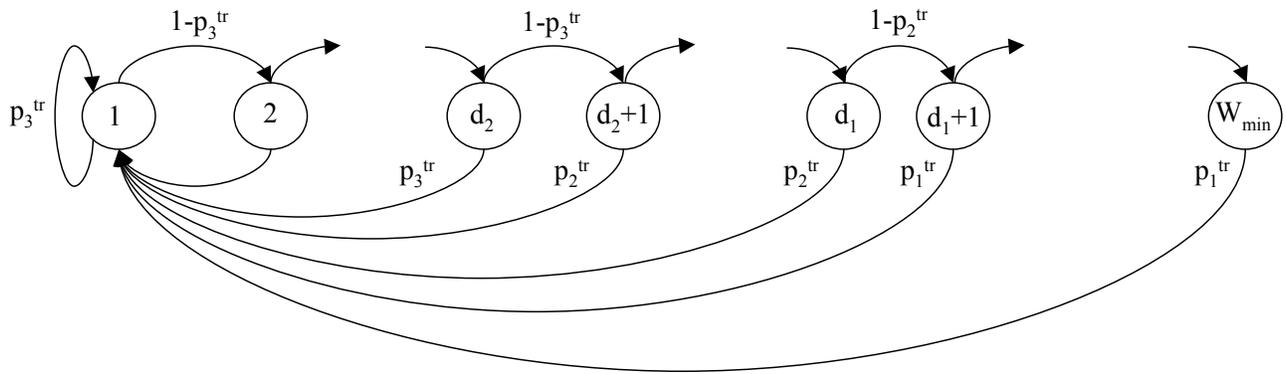,height=17
cm,angle=-90}} \caption[] {\label{fig:unsat_AIFSMC} Transition
through backoff slots in different contention zones for the
example given in Fig.\ref{fig:unsat_contzones}.}
\end{figure}

\clearpage
\begin{figure}[t]
\centering \includegraphics[width = 1.0\linewidth]{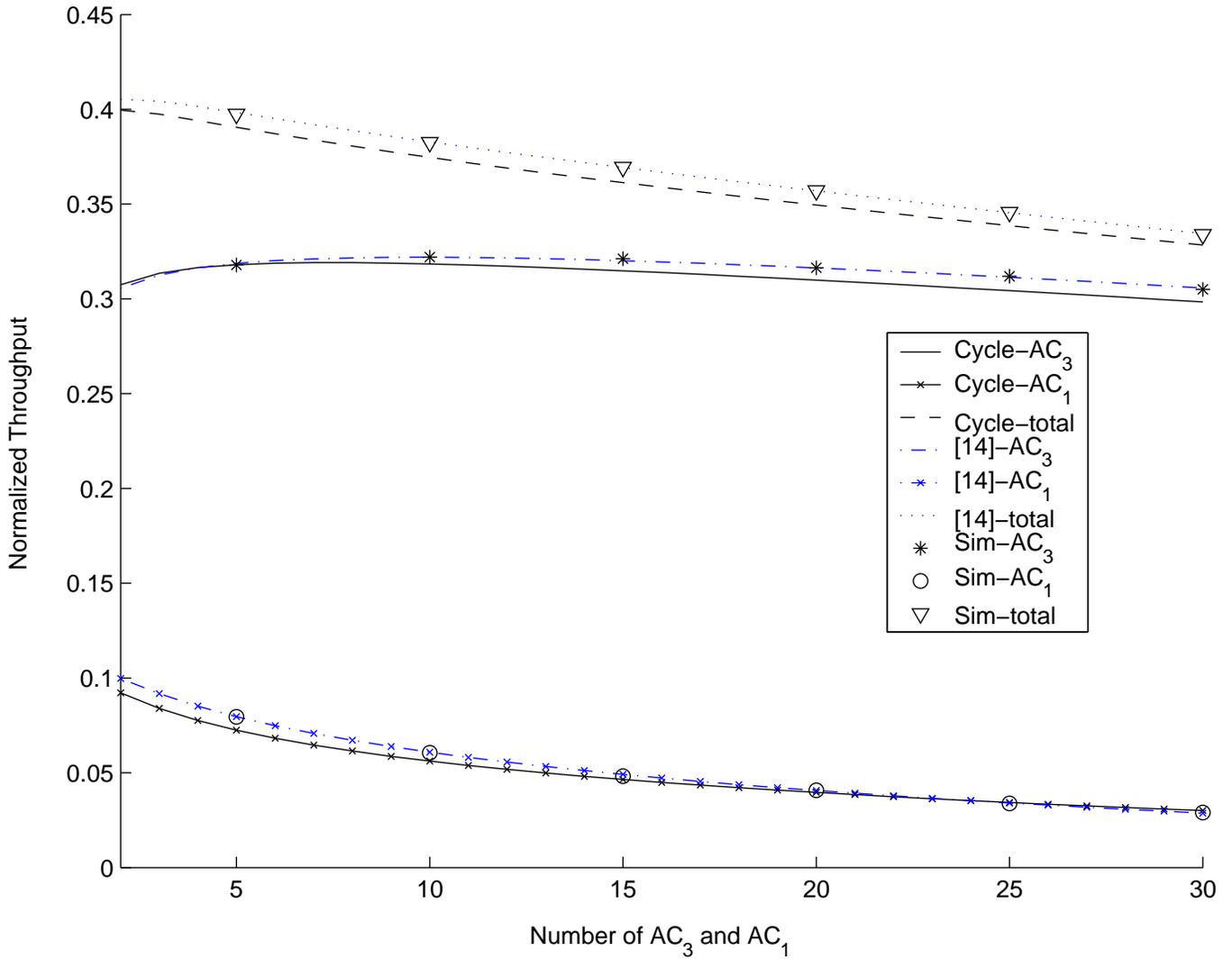}
\caption{Analyzed and simulated normalized throughput of each AC
when both $N_{1}$ and $N_{3}$ are varied from 5 to 30 and equal to
each other for the cycle time analysis. Analytical results of the
model proposed in \cite{Inan07_ICC} are also added for
comparison.} \label{fig:A1_thp_GC07}
\end{figure}

\clearpage
\begin{figure}[t]
\centering \includegraphics[width = 1.0\linewidth]{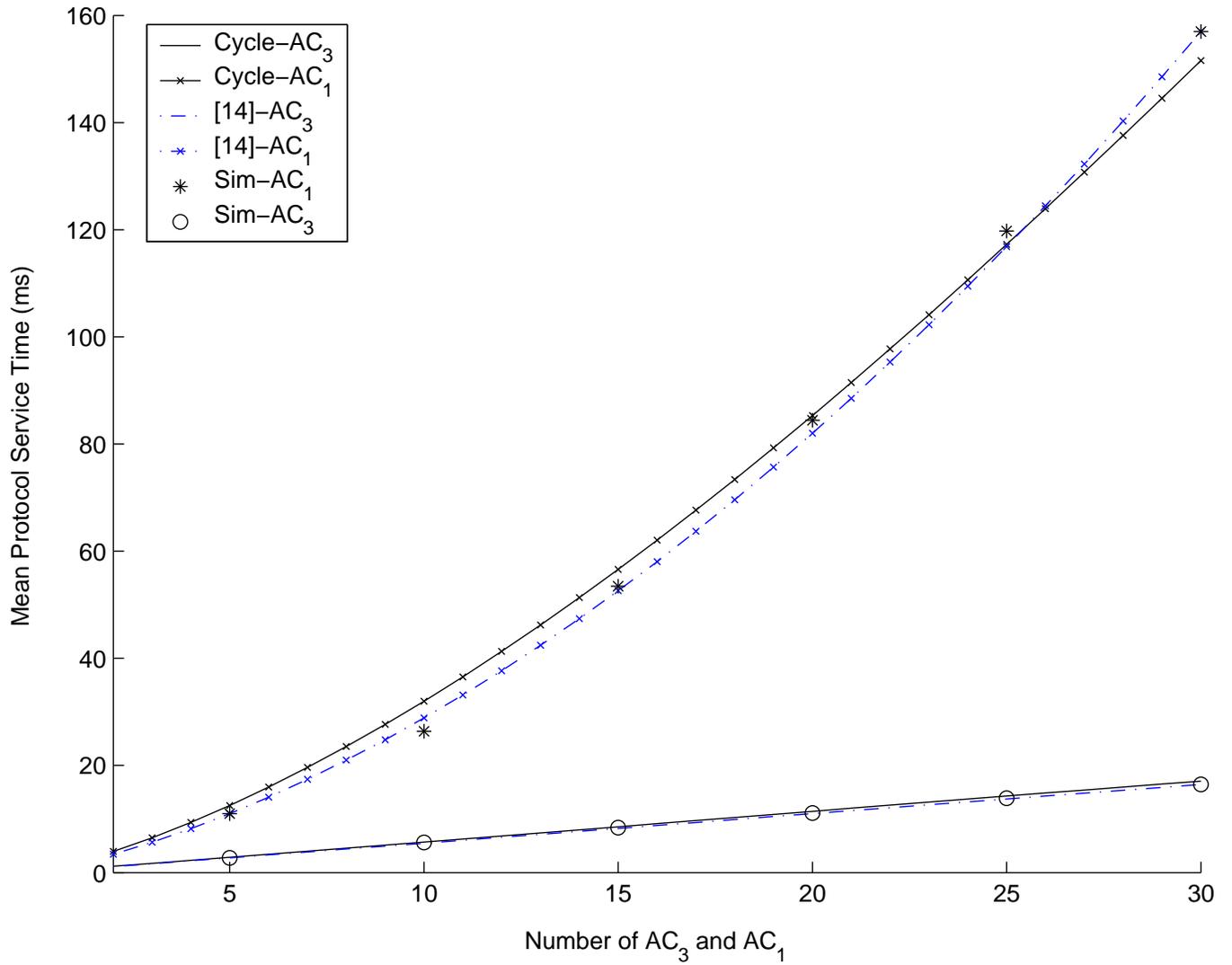}
\caption{Analyzed and simulated mean protocol service time of each
AC when both $N_{1}$ and $N_{3}$ are varied from 5 to 30 and equal
to each other for the proposed cycle time analysis and the model
in \cite{Inan07_ICC}.} \label{fig:A1_mpst_GC07}
\end{figure}


\clearpage
\begin{figure}[t]
\centering \includegraphics[width =
1.0\linewidth]{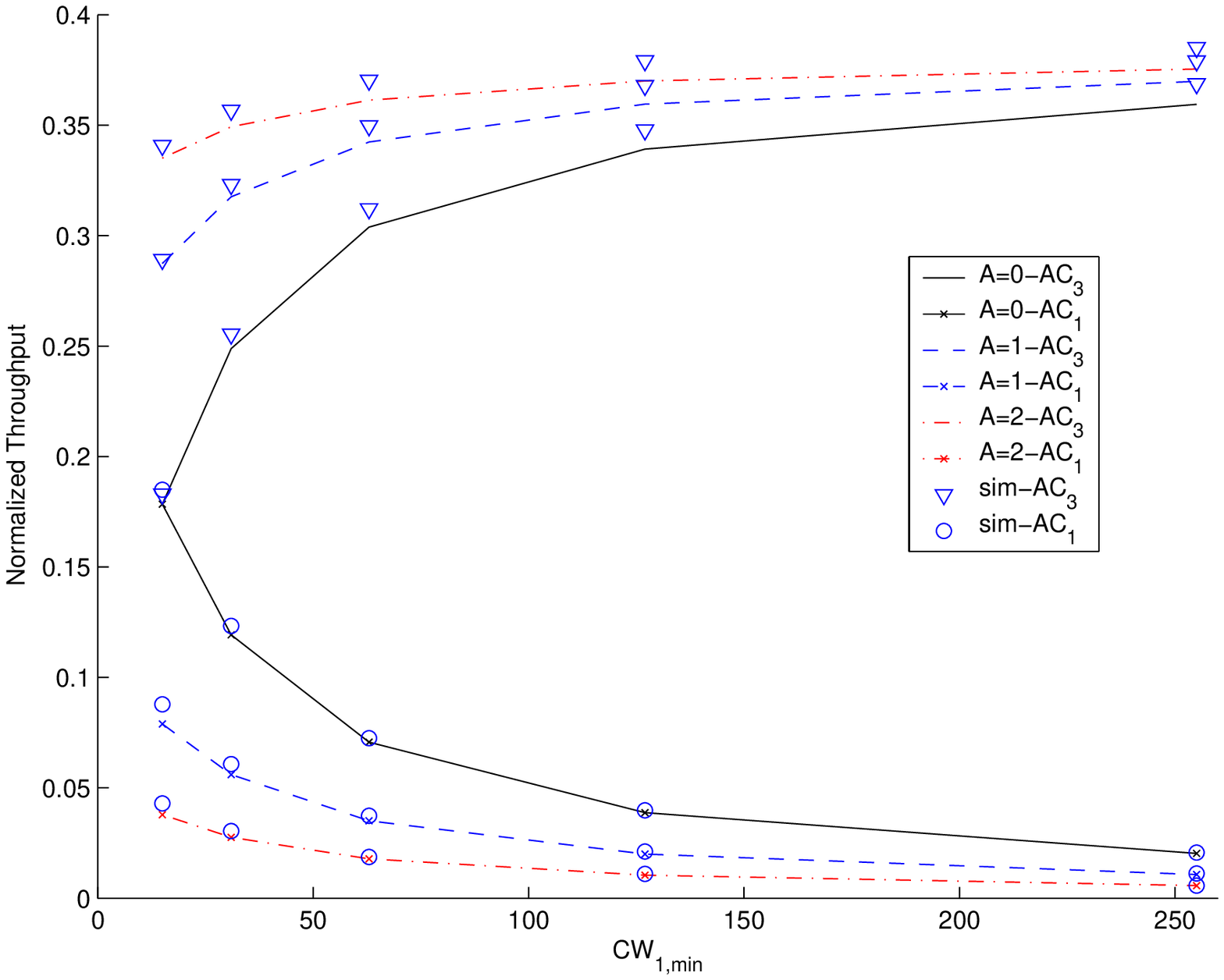} \caption{Analytically calculated
and simulated performance of each AC when $AIFSN_{3}=2$,
$CW_{3,min}=15$, $N_{1}=N_{3}=10$, $AIFSN_{1}$ varies from 2 to 4,
and $CW_{1,min}$ takes values from the set $\{15,31,63,127,255\}$.
Note that $AIFSN_{1}-AIFSN_{3}$ is denoted by $A$.}
\label{fig:v_aifs_cw_1_thp_GC07}
\end{figure}

\clearpage
\begin{figure}[t]
\centering \includegraphics[width =
1.0\linewidth]{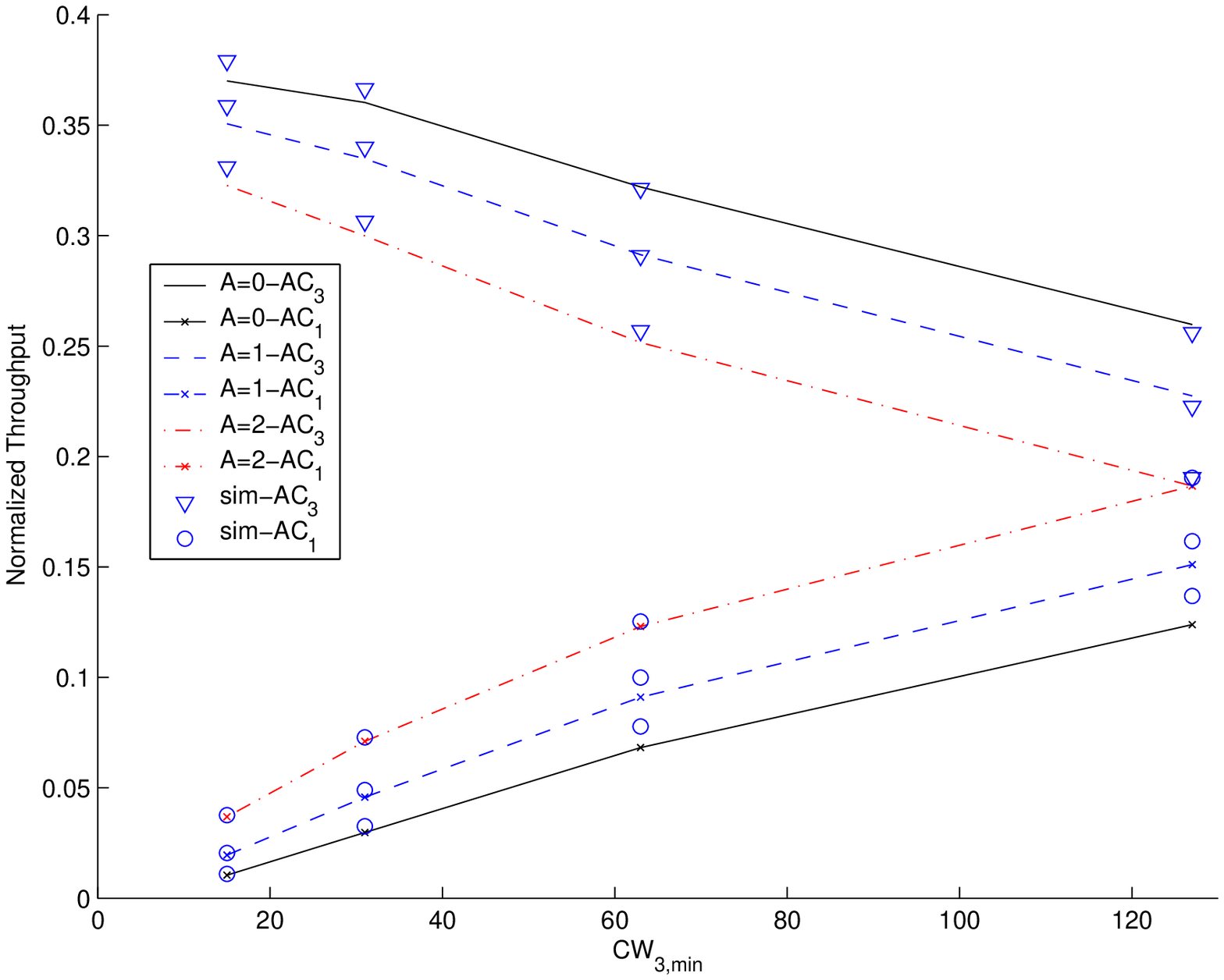} \caption{Analytically calculated
and simulated performance of each AC when $AIFSN_{1}=4$,
$CW_{1,min}=127$, $N_{1}=N_{3}=10$, $AIFSN_{3}$ varies from 2 to
4, and $CW_{3,min}$ takes values from the set $\{15,31,63,127\}$.
Note that $AIFSN_{1}-AIFSN_{3}$ is denoted by $A$.}
\label{fig:v_aifs_cw_3_thp_GC07}
\end{figure}

\clearpage
\begin{figure}[t]
\centering \includegraphics[width = 1.0\linewidth]{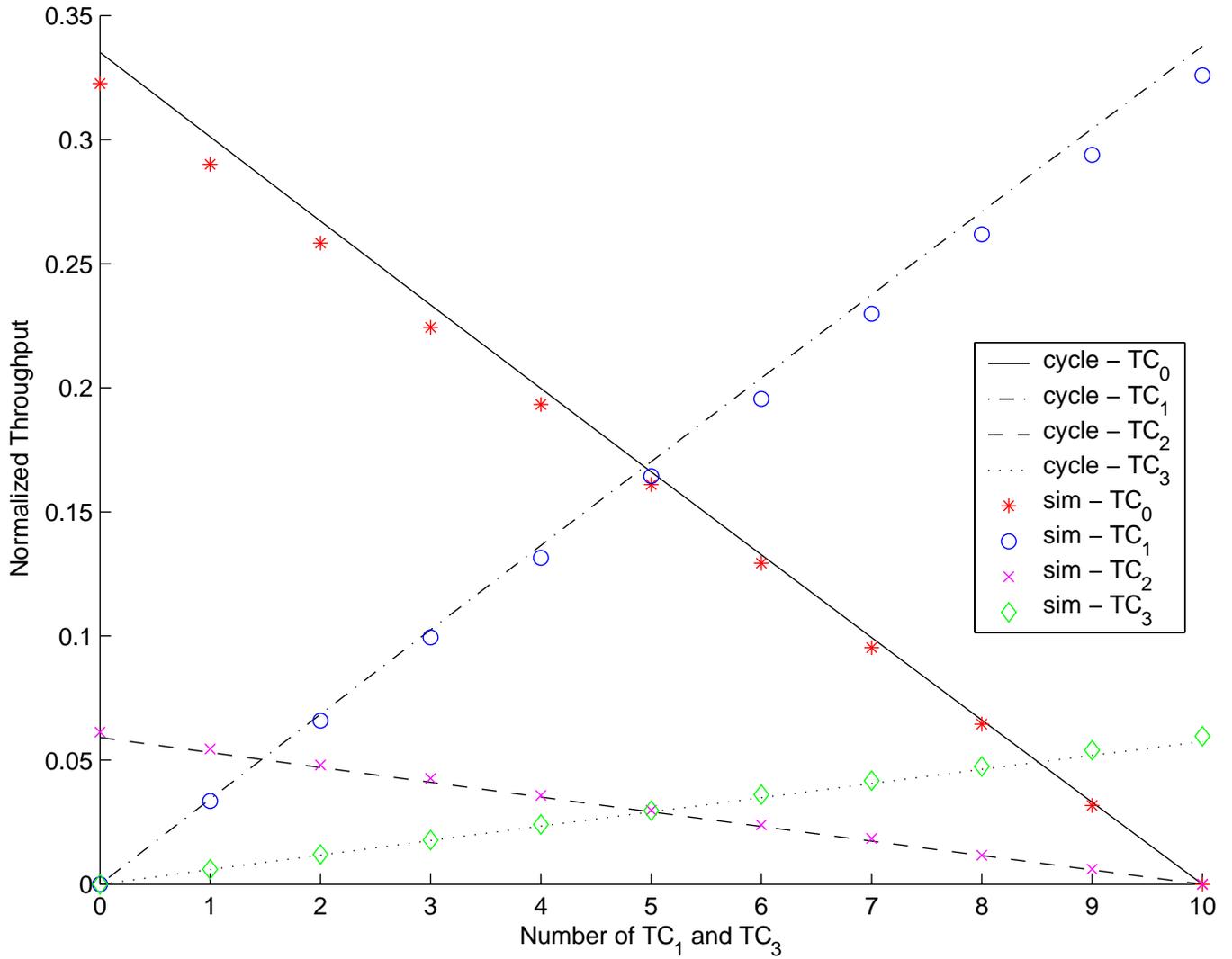}
\caption{Analytically calculated and simulated performance of each
TC when the number of TC$_{1}$ and TC$_{3}$ is varied from 0 to 10
(therefore, TC$_{0}$ and TC$_{2}$ vary from 10 to 0).}
\label{fig:multipleAC_TC}
\end{figure}

\clearpage
\begin{figure}[t]
\centering \includegraphics[width =
1.0\linewidth]{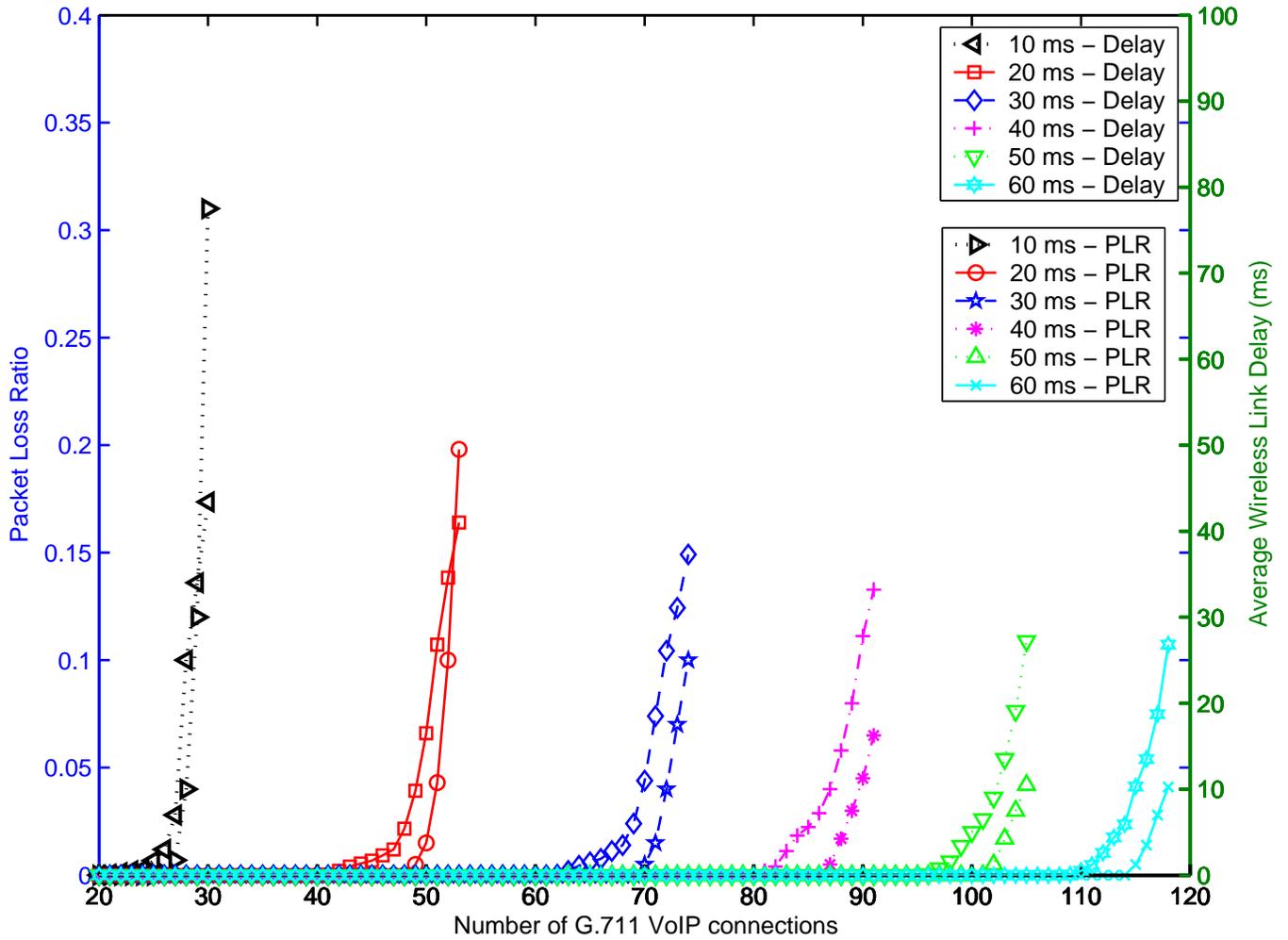} \caption{Packet loss ratio and
average delay in the downlink for increasing number G.711 VoIP
connections.} \label{fig:simG711_plr_delay}
\end{figure}
\clearpage
\begin{figure}
\center {
\begin{tabular}{cc}
\epsfig{file=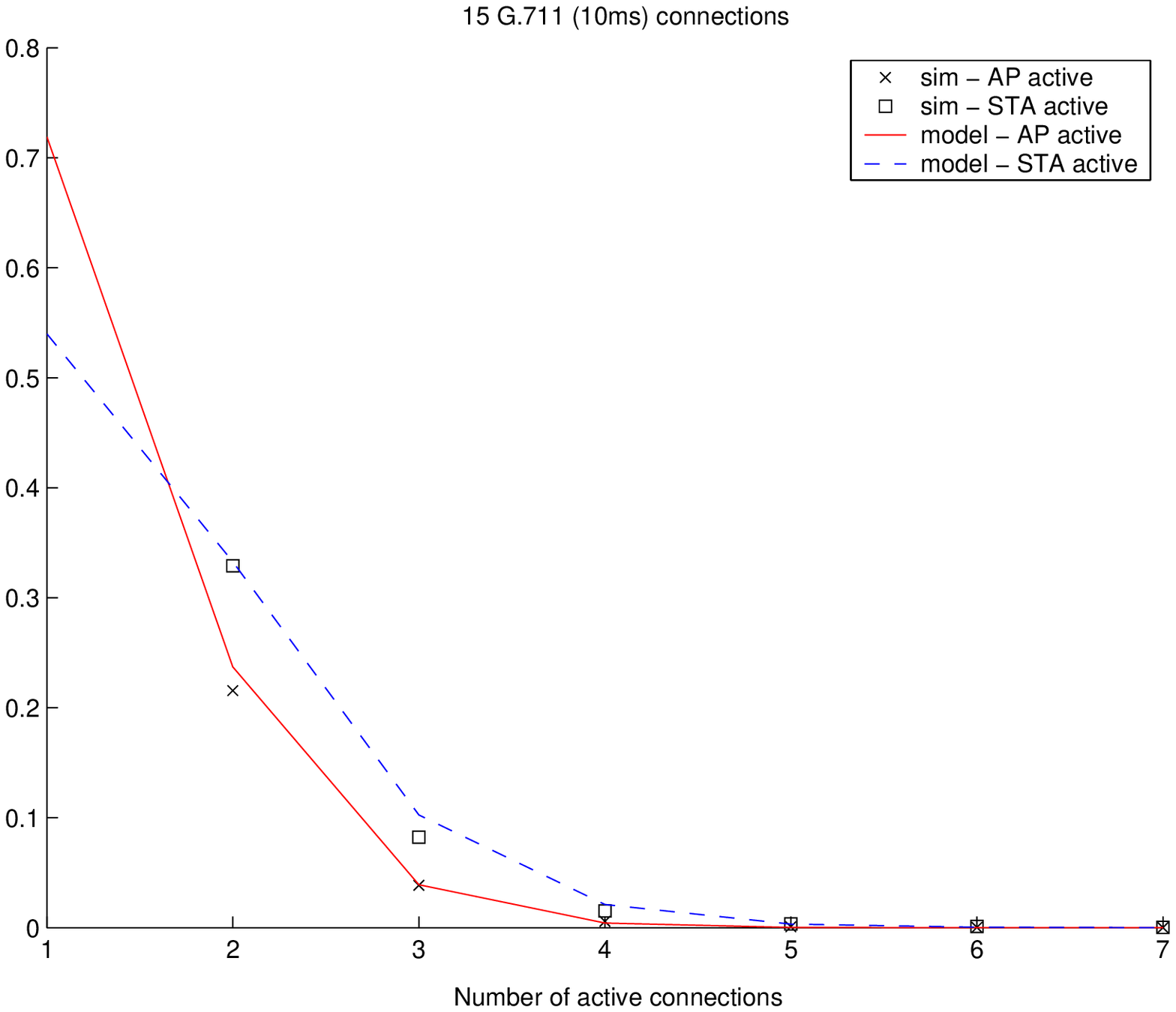,height=7 cm,angle=0}
 &
\epsfig{file=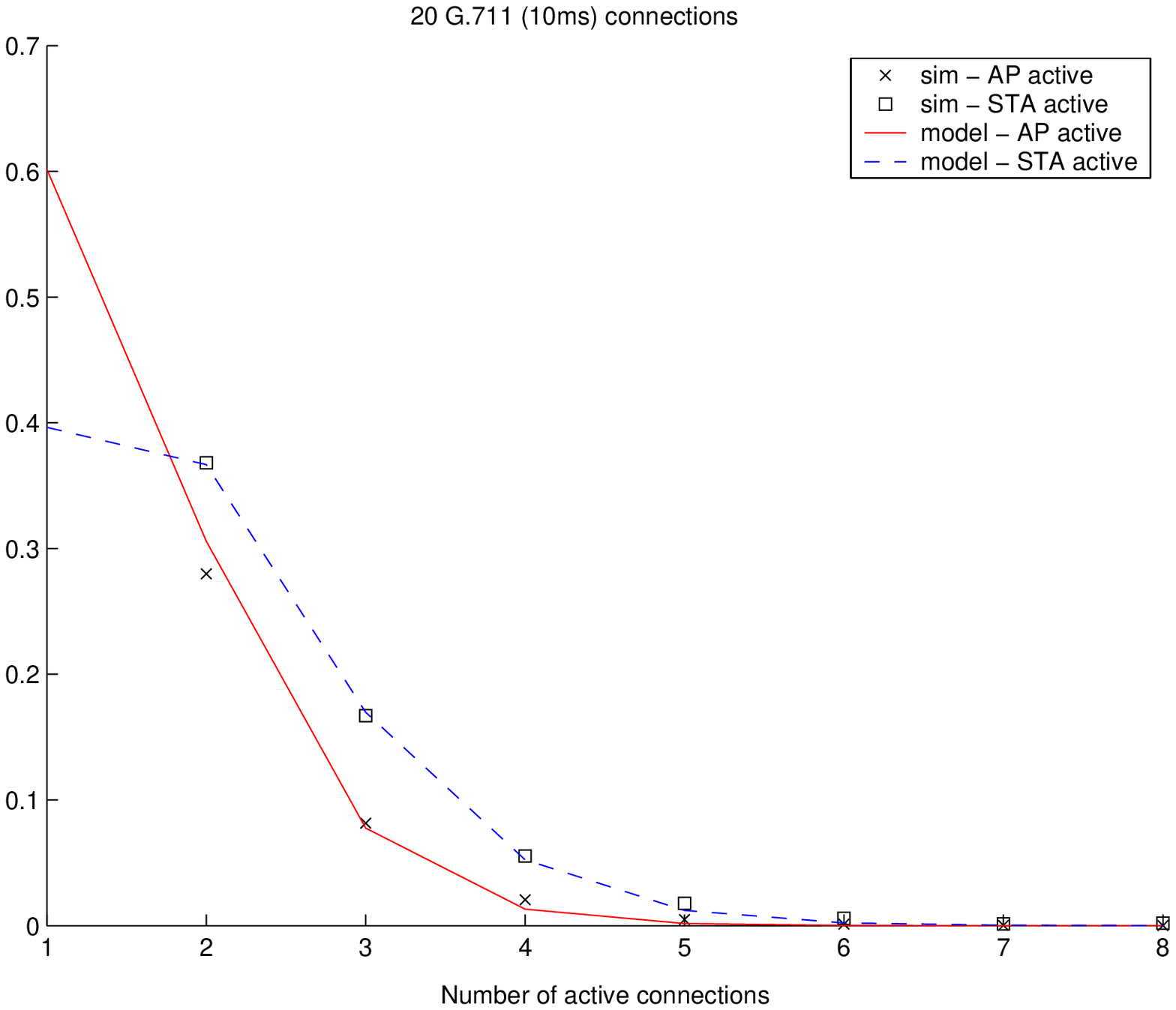,height=7 cm,angle=0} \\
(a) & (b)
\\
\epsfig{file=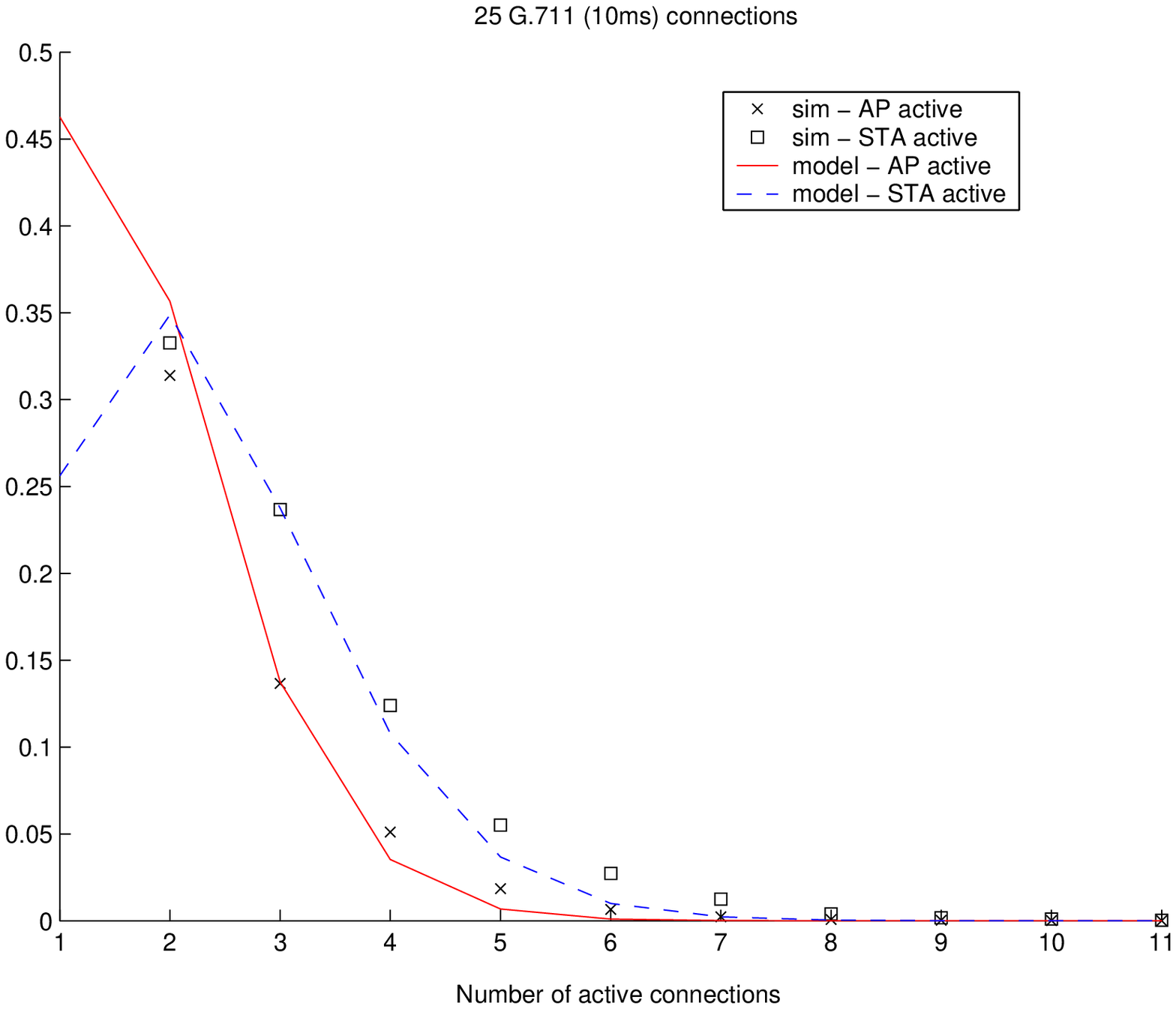,height=7 cm,angle=0} &
\epsfig{file=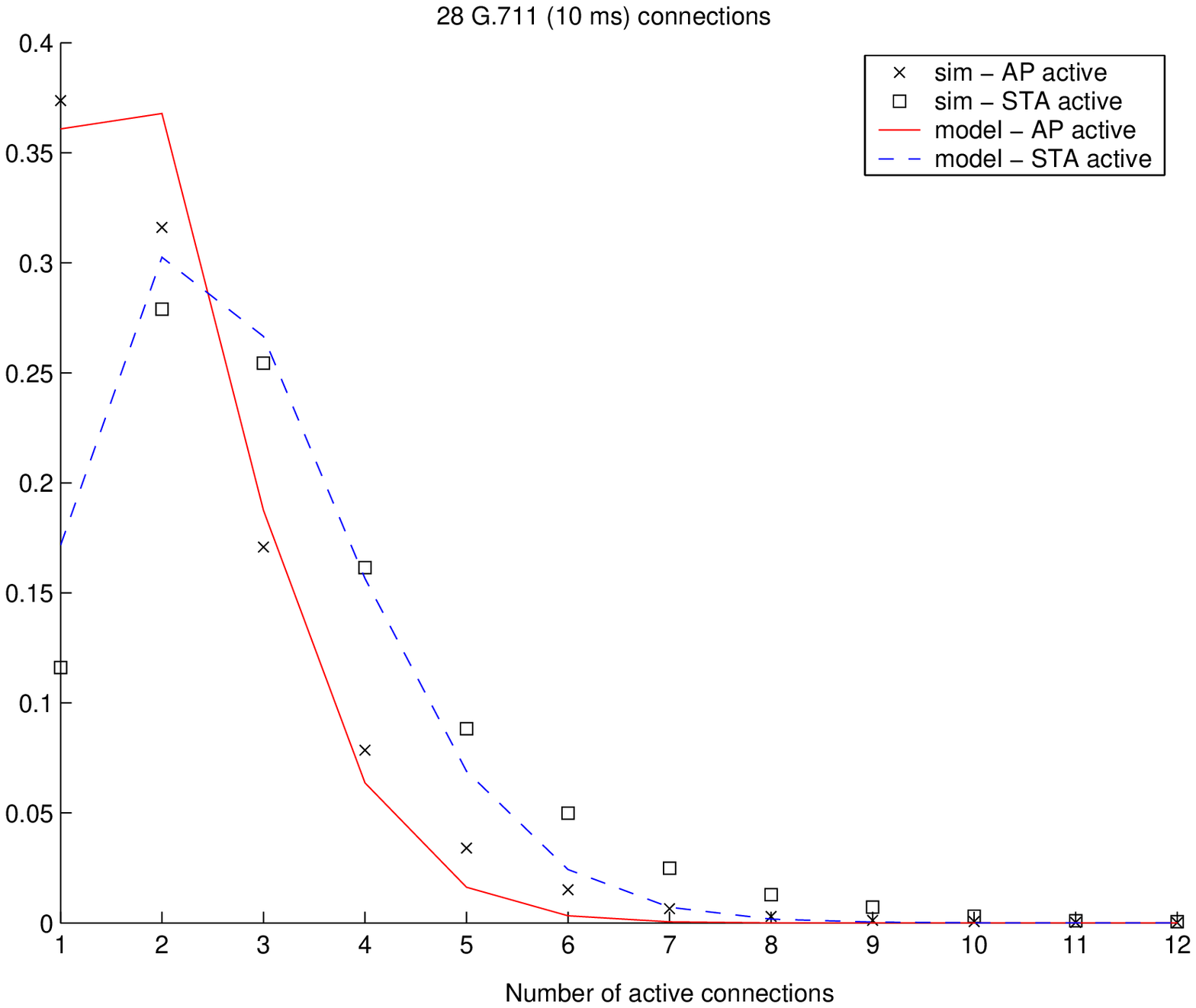,height=7 cm,angle=0} \\
(c) & (d)
\\
\end{tabular}
} \caption{The pdf of active number of TCs given that the TC at
the AP or at the station (denoted as STA) is active in a scenario
consisting of G.711 VoIP connections (10 ms packet intervals). (a)
15 connections. (b) 20 connections. (c) 25 connections. (d) 28
connections. Note that the figures do not present the whole x-axis
(activity profile for large number of stations) for better clarity
on the comparison of simulation and analysis (especially when the
activity probability is not close to zero).}
\label{fig:activity_pdf}
\end{figure}

\clearpage
\begin{table}[t]
\caption{\label{tab:VoIPcapacity} Comparison of the Maximum Number
of VoIP Connections}
\begin{center}
\begin{tabular}{||c|c|c|c|c||}
\hline\multirow{2}{27mm}{Sample Period} & \multicolumn{2}{c|}{G.711} & \multicolumn{2}{c||}{G.729} \\
\cline{2-5} & Analysis/Simulation & \cite{Cai06} &
Analysis/Simulation & \cite{Cai06} \\ \hline 10 ms & 27/27 & 21 & 29/29 & 22 \\
\hline 20 ms & 49/49 & 38 & 56/56 & 43 \\
\hline 30 ms & 70/70 & 53 & 85/85 & 65 \\
\hline 40 ms & 87/87 & 67 & 112/112 & 85 \\
\hline 50 ms & 102/102 & 79 & 139/139 & 106 \\
\hline 60 ms & 115/115 & 89 & 166/166 & 128 \\
\hline
\end{tabular}
\end{center}
\end{table}

\clearpage
\begin{table}[b]
\caption{\label{tab:VoIPdatacapacity} Comparison of the Maximum
Number of G.711 VoIP Connections (Analysis/Simulation) when heavy
background traffic coexist.}
\begin{center}
\begin{tabular}{||p{22mm}|p{26mm}|p{15mm}|p{15mm}|p{15mm}|p{15mm}|p{15mm}|p{15mm}||}
\hline\multirow{2}{*}{} VoIP Codec & Sample Period & \multicolumn{6}{c||}{Number of co-existing two-way background data connections} \\
\cline{3-8}& & 5 & 10 & 15 & 20 & 25 & 30 \\
\hline \multirow{6}{27mm}{G.711}
& 10 ms & 19/18 & 16/16 & 14/14 & 12/12 & 11/11 & 10/10 \\
\cline{2-8} & 20 ms & 35/35 & 29/29 & 26/25 & 23/22 & 21/20 & 19/18 \\
\cline{2-8} & 30 ms & 49/47 & 41/41 & 36/36 & 32/32 & 29/29 & 27/27 \\
\cline{2-8} & 40 ms & 62/62 & 52/52 & 45/45 & 40/40 & 37/37 & 34/34 \\
\cline{2-8} & 50 ms & 73/73 & 61/61 & 53/53 & 47/47 & 43/43 & 40/40 \\
\cline{2-8} & 60 ms & 83/84 & 69/69 & 60/60 & 54/54 & 49/50 & 45/47 \\
\hline
\end{tabular}
\end{center}
\end{table}

\clearpage
\begin{table}[b]
\caption{\label{tab:VoIPvideocapacity} Comparison of the Maximum
Number of Video Connections (Analysis/Simulation) when VoIP flows
coexist.}
\begin{center}
\begin{tabular}{||p{27mm}|p{20mm}|p{15mm}|p{15mm}|p{15mm}|p{15mm}|p{15mm}|p{15mm}||}
\hline\multirow{2}{*}{} & & \multicolumn{6}{c||}{Number of existing two-way G.711 (20 ms) connections} \\
\cline{3-8} & & 5 & 10 & 15 & 20 & 25 & 30 \\ \hline \multirow{3}{27mm}{Number of admitted MPEG-4 flows} & Downlink & 109/110 & 98/100 & 87/88 & 76/78 & 64/65 & 52/54 \\
\cline{2-8} & Uplink & 88/89 & 67/69 & 57/57 & 48/48 & 37/37 & 28/28 \\
\cline{2-8} & Two-way & 54/55 & 46/47 & 41/41 & 34/34 & 28/28 & 19/19 \\
\hline
\end{tabular}
\end{center}
\end{table}